\documentclass[journal,10pt]{IEEEtran}
\usepackage{amsmath,amssymb,amsfonts}
\usepackage{adjustbox}
\usepackage{algorithm,algorithmicx}
\usepackage{algpseudocode}
\usepackage{booktabs}
\usepackage{multirow}
\usepackage{graphicx}
\ifCLASSOPTIONcompsoc
\usepackage[caption=false, font=normalsize, labelfont=sf, textfont=sf]{subfig}
\else
\usepackage[caption=false, font=footnotesize]{subfig}
\fi
\usepackage{threeparttable}

\begin{document}

\title{An Identity-based Batch Verification Scheme for VANETs Based on Ring Signature with Efficient Revocation}

\author{Feng Liu, Qi Wang

\thanks{This paper was presented in part at the 2019 IEEE Vehicular Networking Conference held on December 4-6, 2019 in Los Angeles, California.}
\thanks{F. Liu and Q. Wang are with the Department of Computer Science and Engineering, Southern University of Science and Technology, Shenzhen, Guangdong 518055, China (e-mail: liuf2017@mail.sustech.edu.cn; wangqi@sustech.edu.cn)}

}

{}

\maketitle

\begin{abstract}
Vehicular ad-hoc networks (VANETs) are one of the most important components in Intelligent Transportation System (ITS), which aims to provide secure and efficient communication between vehicles. Safety-critical vehicular communication requires security, privacy, and auditability. To satisfy these requirements simultaneously, several conditional privacy-preserving authentication schemes are proposed by employing ring signatures. However, these methods have paid little attention to the issues like \textit{how to choose the valid ring members} or \textit{how to set up a ring}. In this paper, we introduce an efficient conditional privacy-preserving scheme which provides an appropriate approach establishing the list of ring members with efficient revocation. Moreover, our proposed scheme also provides batch verification to significantly reduce the computational cost. According to the analysis of security, our scheme is sufficiently resistant against several common attacks in VANETs. The performance results show that the proposed scheme is efficient and practical with both low computation and communication cost.
\end{abstract}

\begin{IEEEkeywords}
VANETs, ring signature, conditional privacy, batch verification
\end{IEEEkeywords}

\IEEEpeerreviewmaketitle

\section{Introduction}\label{I}
Vehicular Ad-hoc Networks (VANETs), as a special kind of Mobile Ad-hoc Networks (MANETs) optimized for vehicular environments, plays an important role in Intelligent Transportation Systems (ITS). In a typical scenario of ITS, each vehicle broadcasts traffic-related information, such as its speed, position, road condition, and others through VANETs. After receiving these broadcast messages, vehicles can analyze and extract meaningful information to drivers, or take corresponding control actions in emergencies. In this way, the applications of VANETs reduce the rate of traffic accident, and thereby road safety and efficiency will be greatly enhanced. Besides, this technology is also meaningful to automated vehicles because VANETs make it possible for vehicles to communicate with each other. However, due to the high demand for road safety features, to design a practical protocol for VANETs is highly nontrivial. Numerous proposed schemes are built based on the IEEE 802.11p standard.

\begin{figure}[htbp]
    \centering
    \includegraphics[width=\linewidth]{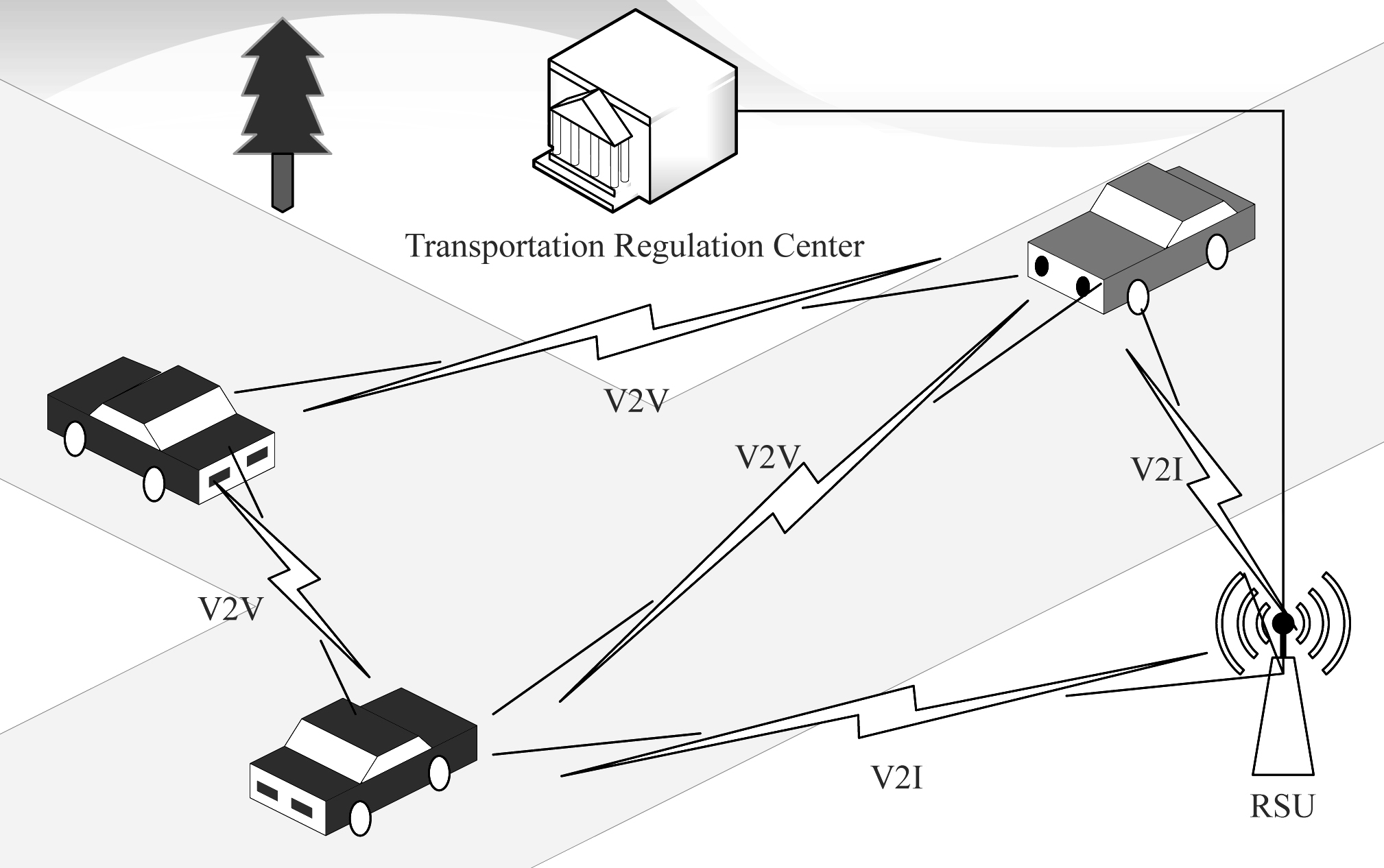}
    \caption{A typical scenario of VANETs}
    \label{fig:overall}
\end{figure}

In IEEE 802.11p, the participants on the road are classified into two categories, i.e., On-Board Units (OBUs) and Road-Side Units (RSUs). Typically, each vehicle is equipped with an OBU for broadcasting messages and handling the received messages.
The RSUs are usually fixed along roads as the base stations to provide Internet access and extra road-related information for vehicles.
Therefore, VANETs provide two different types of communication, namely, Vehicle-to-Infrastructure (V2I) and Vehicle-to-Vehicle (V2V) as shown in Figure~\ref{fig:overall}. As a result, drivers can be reminded of road conditions by receiving the broadcast messages from other vehicles or RSUs in VANETs.
In practice, a trusted party, called Transportation Regulation Center (TRC), is needed to administrate the whole network. For example, RSUs can connect with TRC for obtaining extra traffic information.

However, due to the open environment of VANETs, an attacker could send a forged message to confuse nearby drivers, which may further cause potential traffic hazards. To maintain road safety, it is essential to authenticate the validity of a message. The first serious discussion of road safety was proposed in 2002 and digital signature was utilized therein~\cite{el2002security}. After that, a considerable amount of work has been proposed based on different digital signature schemes~\cite{petit2014pseudonym}.

When applying various digital signature schemes in VANETs, each message is attached with some extra information including signature, signer's certificate and so on. The extra information may be used to link to the driver's true identity, and even lead to privacy disclosure. Therefore, \textit{how to keep the anonymity of senders} while the message can be authenticated by verifiers becomes another essential issue in VANETs. A common approach is to replace the true identity with a random-like string called pseudonym. On the other hand, in some specific scenarios such as in a traffic accident, the true identity of senders should be revealed by law enforcement. To this end, a function of auditability should be provided. Thus, authentication, anonymity, and auditability are three basic requirements when designing a feasible scheme for communication in VANETs. Besides, due to the limited computation and storage capability of both OBUs and RSUs, efficiency should also be considered in VANETs.

\subsection{Our Contribution}

In this paper, we propose a hybrid scheme for VANETs by taking advantage of ring signature, identity-based cryptography (IBC) and symmetric cryptography. Since ring signature allows a signer to create signatures on behalf of an ad-hoc group without any additional setup, and achieves unconditional anonymity for the signer, it seems quite suitable to put ring signature into ad-hoc networks like VANETs. Unfortunately, no setup means any vehicle (even a malicious one) can generate a signature on behalf of a set of vehicles, and unconditional annoymity infers that there lacks an efficient approach to trace and revoke a vehicle (i.e., lack of auditability). These two issues make it difficult to apply ring signature in VANETs.

To address these issues, we suggest utilizing RSUs to distribute the group or ring information to valid vehicles. For clarity, we divide the auditability into two aspects: pseudonym resolution and revocation.
In terms of pseudonym resolution, we introduce a new entity called {\em Law Enforcement Authority} (LEA) to reveal the signer's identity. As for pseudonym revocation, we incorporate the KUNodes algorithm~\cite{boldyreva2008identity} into the proposed scheme so that the size of key updates decreases from linear down to logarithmic. More precisely, our contributions are summarized in the following:

\begin{itemize} 
 \item We propose a novel scheme for VANETs based on identity-based ring signature, where the procedure of creating a ring is restricted by RSUs.

 \item We provide a batch mode to accelerate message verification in VANETs. As indicated by performance, this makes our scheme highly efficient. To the best of our knowledge, this is the first attempt that applies ring batch verification in VANETs.
 
 \item We present an efficient way to make vehicles auditable by using different technologies including general one-way hash functions and KUNodes algorithm.

 \item We implement the scheme in the Raspberry Pi 4 Model B platform and give a comprehensive analysis based on the platform.
\end{itemize}

\subsection{Related Work}~\label{relatedwork}

Numerous studies have attempted to employ pseudonym schemes to assure authentication and privacy simultaneously. Generally, the cryptographic tools utilized include public key infrastructure (PKI), identity-based cryptography (IBC), group signature, ring signature and so on (for recent surveys, see~\cite{petit2014pseudonym,ali2019authentication,sharma2019survey}).

At the early stage of the study, PKI was most widely used in VANETs.
In these schemes based on PKI, Certificate Authority (CA) is needed as a trusted party. Each vehicle broadcasts messages attached to the corresponding signatures and public-key certificates. Taking the SeVeCom project~\cite{wiedersheim2009sevecom} as an example, the elliptic curve digital signature algorithm (ECDSA) is utilized to assure efficiency. The pseudonym used in VANETs is composed of two parts: a short-term key and its corresponding certificate. Since the transmission of certificates will increase the communication overhead, it was alternatively suggested to employ IBC instead of PKI~\cite{el2002security,kamat2006identity}.

Similarly, IBC-based schemes also adopt a set of short-term public keys to form vehicles' pseudonyms, while the procedure of pseudonym issuance differs. Note that in IBC-based schemes, a new entity called private key generator (PKG) is introduced to replace CA. Thus, certificates are not attached when broadcasting messages in these schemes, and the communication overhead is thereby decreased.

However, pseudonym changing becomes a major problem in both PKI-based and IBC-based pseudonym schemes, as only using a single pseudonym is not sufficient to preserve vehicles' privacy. In a simple setting, each vehicle is equipped with a set of public keys, each of which can be viewed as an unlinkable pseudonym, and will expire after a fixed amount of time. Wiedersheim et al.~\cite{wiedersheim2010privacy} pointed out that simple pseudonym change is not enough to preserve
privacy. There have been attempts on the strategy of pseudonym change, e.g., mix-zone-based~\cite{ying2013dynamic} and mix-context-based~\cite{gerlach2007privacy}. However, it is still mysterious to formalize the relationship between pseudonym change strategies and privacy level~\cite{petit2014pseudonym,boualouache2017survey}.

The issue of pseudonym change can be eliminated in the schemes based on group signature and ring signature~\cite{petit2014pseudonym,calandriello2007efficient}. In these schemes, messages are signed under the identity of a certain group rather than a single vehicle's pseudonym.  Group signature-based schemes allow a vehicle to sign a message anonymously on behalf of the group. In group signatures, a special entity called group manager can reveal any signer's real
identity from the corresponding signature. Till now, there has been little agreement on the choice of group manager~\cite{petit2014pseudonym}. As the administrator of the group, the group manager has the privilege to add or delete a group member. It is then straightforward to achieve auditability of group members by the group manager. Some researchers~\cite{park2011rsu,zhang2009scalable} suggested that RSUs serve as group managers. However, RSUs are vulnerable to some extent, and this setting is not sufficient to guarantee group members' privacy.

\begin{figure}[htbp]
   \centering
   \includegraphics[width=\linewidth]{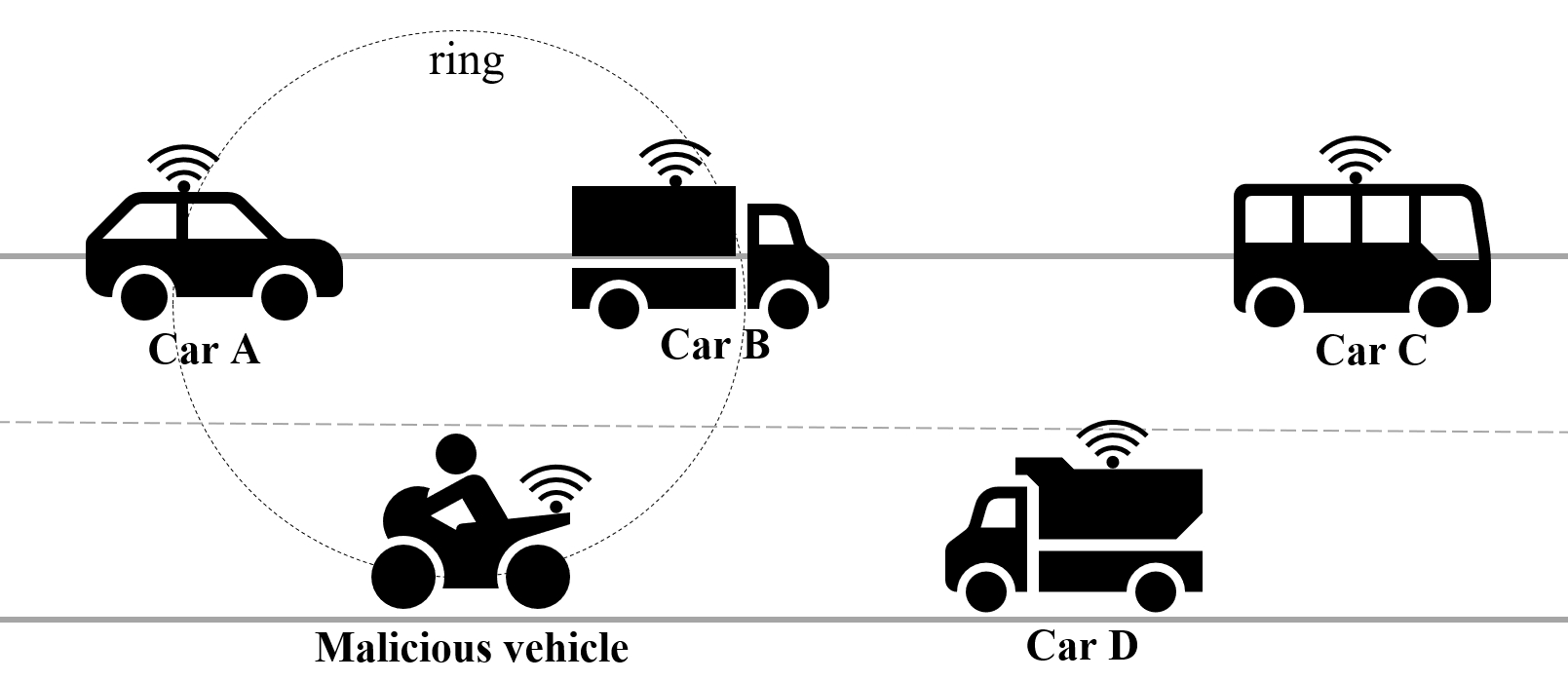}
   \caption{A negative case in ring signature-based schemes}
   \label{fig:0t}
\end{figure}

In comparison, ring signature-based schemes~\cite{zeng2015privacy,jiang2014anonymous,zeng2018concurrently} further remove group managers by involving a set of different vehicles' public keys as a ring. In existing ring signature-based schemes, e.g.,~\cite{zeng2015privacy}, each vehicle can collect other vehicles' public keys on the road and thereby checks the validity of ring members before verification. Unfortunately, this would lead to verification failure when a
malicious vehicle broadcasts an invalid public key. As depicted in Figure~\ref{fig:0t}, due to the existence of a malicious vehicle in the ring, the generated signature by the whole ring would be rejected by other vehicles. In existing such schemes, privacy is the main concerning issue while the lack of auditiability of ring members for VANETs is still a problem~\cite{chaurasia2011conditional,zeng2015privacy}.

Petit et al.~\cite{petit2014pseudonym} emphasized that these categories are not hard-edged so that several recent works combined different techniques from the previous categories. Survey~\cite{ali2019authentication} listed these hybrid schemes and discussed their security and efficiency in performance. These results pointed out that recently proposed schemes attempted to apply the batch verification of signatures into the verification procedure, which can greatly reduce the computation cost comparing with single verification.

As we mentioned above, to achieve auditability in VANETs, pseudonym resolution and pseudonym revocation should be taken into account. A common method to revoke users in PKI setting is adopting certificate revocation lists (CRLs). Since the size of CRLs increases linearly with the number of revoked vehicles, this method seems not practical in VANETs because the capacity of both OBUs and RSUs is limited~\cite{ali2019authentication}. To cope with the problem, Khodaei and
Papadimitratos~\cite{Khodaei_2020} proposed a solution by splitting CRLs into CRL pieces and using Bloom Filter to compress these pieces. On the other hand, Bloom Filter is a probabilistic data structure in which false positive may occur. As for the schemes based on group signature, it is straightforward to achieve auditability by the group manager. But in those schemes based on IBC and ring signature, there lacks native support on the revocation of an identity. Thus, it is necessary to find a suitable approach to achieve auditability for VANETs.

\subsection{Outline}

The remainder of this paper is organized as follows. Section~\ref{II} introduces some preliminary cryptographic primitives. The system model of our proposed scheme is illustrated in Section~\ref{III}. In Section~\ref{IV}, a description of our schemes is given in detail. After that, security analysis and performance analysis are provided in Section~\ref{V} and Section~\ref{VI}, respectively. Finally, Section~\ref{VII} concludes this paper and proposes some potential future work.

Note that part of this work was presented at the 2019 IEEE Vehicular Networking Conference (VNC'19), December 4-6, 2019, Los Angeles, California~\cite{liu2019ibrs}. In this journal version, we present the comprehensive framework with efficient revocation mechanism in detail and conduct more in-depth experiments and analysis of the proposed scheme.

\section{Preliminaries}\label{II}

In this section, we briefly introduce the cryptographic primitives involved in the proposed scheme, including bilinear pairings, identity-based encryption, and identity-based ring signature.

\subsection{Bilinear pairings}\label{II-B}

Bilinear pairings have been widely used to design various cryptographic schemes over the last two decades~\cite{boneh2001identity,chow2005efficient,zhang2002id}.
Let $\mathbb{G}_1$, $\mathbb{G}_2$, $\mathbb{G}_T$ be three cyclic groups of the same prime order $q$.
Assume that the discrete logarithm problem in $\mathbb{G}_1$, $\mathbb{G}_2$ and $\mathbb{G}_T$ is hard.
Let $\hat{e}:\mathbb{G}_1 \times \mathbb{G}_2\rightarrow \mathbb{G}_T$ be a bilinear pairing with the following properties:

\begin{itemize}
  \item[1.] Bilinearity: $\forall P\in \mathbb{G}_1$, $\forall Q\in\mathbb{G}_2$ and $\forall a,b\in \mathbb{Z}_q^*$, $\hat{e}(aP,bQ)=\hat{e}(P,Q)^{ab}$;
  \item[2.] Non-degenerateness: $\exists P\in \mathbb{G}_1$, $\exists Q\in \mathbb{G}_2$ such that $\hat{e}(P,Q)\neq1$;
  \item[3.] Computability: $\forall g_1\in \mathbb{G}_1$,$\forall g_2\in \mathbb{G}_2$, there is an efficient algorithm to compute $\hat{e}(g_1,g_2)$.
\end{itemize}

There are two hard-problem assumptions in bilinear pairings, i.e., Computational Bilinear Diffie-Hellman (CBDH) Problem and Decisional Bilinear Diffie-Hellman (DBDH) Problem, as described in the following.
\begin{itemize}
    \item CBDH: Given $P\in\mathbb{G}_1$, $aQ$, $bQ$, $cQ\in(\mathbb{G}_2)^3$, where $a, b, c\in_R(\mathbb{Z}_q^*)^3$, it is difficult to calculate $\hat{e}(P,Q)^{abc}$.
    \item DBDH: Given $P\in\mathbb{G}_1$, $aQ$, $bQ$, $cQ\in(\mathbb{G}_2)^3$, $h\in\mathbb{G}_T$, where $a, b, c\in_R(\mathbb{Z}_q^*)^3$, it is difficult to determine whether or not $h=\hat{e}(P,Q)^{abc}\ \mathsf{mod}\ q$.
\end{itemize}

According to the relation between $\mathbb{G}_1$ and $\mathbb{G}_2$, bilinear pairings can be divided into three types~\cite{galbraith2008pairings}:

\begin{itemize}
    \item Type 1: $\mathbb{G}_1=\mathbb{G}_2$;
    \item Type 2: $\mathbb{G}_1\neq\mathbb{G}_2$, but there is an efficiently computable homomorphism $\phi:\mathbb{G}_2\rightarrow\mathbb{G}_1$;
    \item Type 3: $\mathbb{G}_1\neq\mathbb{G}_2$, and there is no efficiently computable homomorphism between $\mathbb{G}_1$ and $\mathbb{G}_2$.
\end{itemize}

It was shown in~\cite{galbraith2008pairings} that Type 3 is much more suitable in practical applications since it can offer better performance and flexibility than other types under the same security level. But several proposed schemes did not take Type 3 into account due to the lack of homomorphism from $\mathbb{G}_2$ to $\mathbb{G}_1$ in Type 3.

\subsection{Identity-based cryptography}\label{II-C}

Bilinear pairings are usually used to construct identity-based encryption and signature schemes. Compared to traditional PKI, identity-based cryptography avoids CA since each user's public key can be automatically derived from the corresponding identity (e.g., user's phone number, email address) by PKG. In general, a common identity-based cryptosystem contains two basic algorithms.

\begin{itemize}
    \item [1.] $\mathsf{Setup}(1^\kappa)\rightarrow \mathsf{PP}$: Taking the input of security parameter $\kappa$, the algorithm $\mathsf{Setup}(1^\kappa)$ first chooses a master secret key $s\in_R\mathbb{Z}_q^*$, and then outputs the public parameter $\mathsf{PP}=\{\mathbb{G}_1,\mathbb{G}_2,\mathbb{G}_T,P,Q,PK_1,PK_2,q,\hat{e},H_1,H_2\}$, where $\mathbb{G}_1,\mathbb{G}_2,\mathbb{G}_T,q,\hat{e}$ are described in
      Section~\ref{II-B}, $P$ is a generator of $\mathbb{G}_1$, $Q$ is a generator of $\mathbb{G}_2$, and $PK_1=s\cdot P$, $PK_2=s\cdot Q$.
      $H_1$, $H_2$ and $H$ are three cryptographic hash functions where $H_1:\{0,1\}^*\rightarrow\mathbb{G}_1$, $H_2:\{0,1\}^*\rightarrow\mathbb{G}_2$, $H:\{0,1\}^*\rightarrow\mathbb{Z}_q^*$.
    \item [2.] $\mathsf{KeyGen}(\mathsf{ID}_i)\rightarrow \{pk_i,sk_i\}$: When user $i$ wants to obtain public key and private key from the system, first $i$ needs to send the specific identity $\mathsf{ID}_i$ to the system. After authentication is accepted, the system runs $\mathsf{KeyGen}(\mathsf{ID}_i)$ to derive the user's public key $pk_i$ and private key $sk_i$, where $pk_i=H_1(\mathsf{ID_i})$ and $sk_i=s\cdot pk_i$.
\end{itemize}

The first practical identity-based encryption scheme was proposed in 2001 by Boneh and Franklin~\cite{boneh2001identity}. In the rest of this paper, we use $\mathsf{Enc}_{pk_i}(\cdot)$ and $\mathsf{Dec}_{sk_i}(\cdot)$ to denote the variant of identity-based encryption and decryption algorithms in~\cite{boneh2001identity}, respectively.

Furthermore, an identity-based ring signature is also used in our scheme, which requires verifiers to verify messages through a specific set of signers. Note that this is different from the traditional signature. 
More precisely, we adopt CYH identity-based ring signature scheme~\cite{chow2005efficient} in our proposed scheme. Note that the CYH signature introduced here can be replaced by other lightweight identity-based ring signature schemes. There are two key algorithms: signature algorithm $\mathsf{Sign}_{sk_i}(m,L)\rightarrow \sigma$ and verification algorithm $\mathsf{Verify}(m,L,\sigma)\rightarrow 0/1$, where $m$ and $L$ denote the message and the ring-member list, respectively.

\subsection{KUNodes algorithm}~\label{II-D}

A common approach to update a  revocation list efficiently in IBC is KUNodes algorithm, which was first introduced by Blodyreva et al.~\cite{boldyreva2008identity}.

\begin{algorithm}
  \begin{algorithmic}[htbp]
  \State $X,Y\gets \emptyset$
  \For {($v_i,T_i)\in RL$}
    \If {$T_i \leq T$}
      \State add $\mathsf{Path}(v_i)$ to $X$
    \EndIf
  \EndFor
  \For {$x\in X$}
    \If {$x_{left} \notin X$}
      \State add $x_{left}$ to $Y$
    \ElsIf {$x_{right} \notin X$}
      \State add $x_{right}$ to $Y$
    \EndIf
  \EndFor
  \If {$Y=\emptyset$}
    \State add $\mathsf{root}$ to $Y$
  \EndIf
  \State Return $Y$
 \caption{$\mathsf{KUNodes}(\mathsf{BT},RL,T):$}~\label{alg:KUNode}
\end{algorithmic}
\end{algorithm}

In general, this algorithm takes as input a binary tree $\mathsf{BT}$, revocation list $RL$, and a period $T$. We denote by $\mathsf{root}$ the root node. For a non-leaf node $x$, we use $x_{left}$ and $x_{right}$ to represent its left child node and right child node, respectively. For a leaf node $v$, $\mathsf{Path}(v)$ means the set of nodes on the path from $v$ to $\mathsf{root}$. The description of KUNodes algorithm is given in Algorithm~\ref{alg:KUNode}.

\begin{figure}[htbp]
   \centering
   \includegraphics[width=0.8\linewidth]{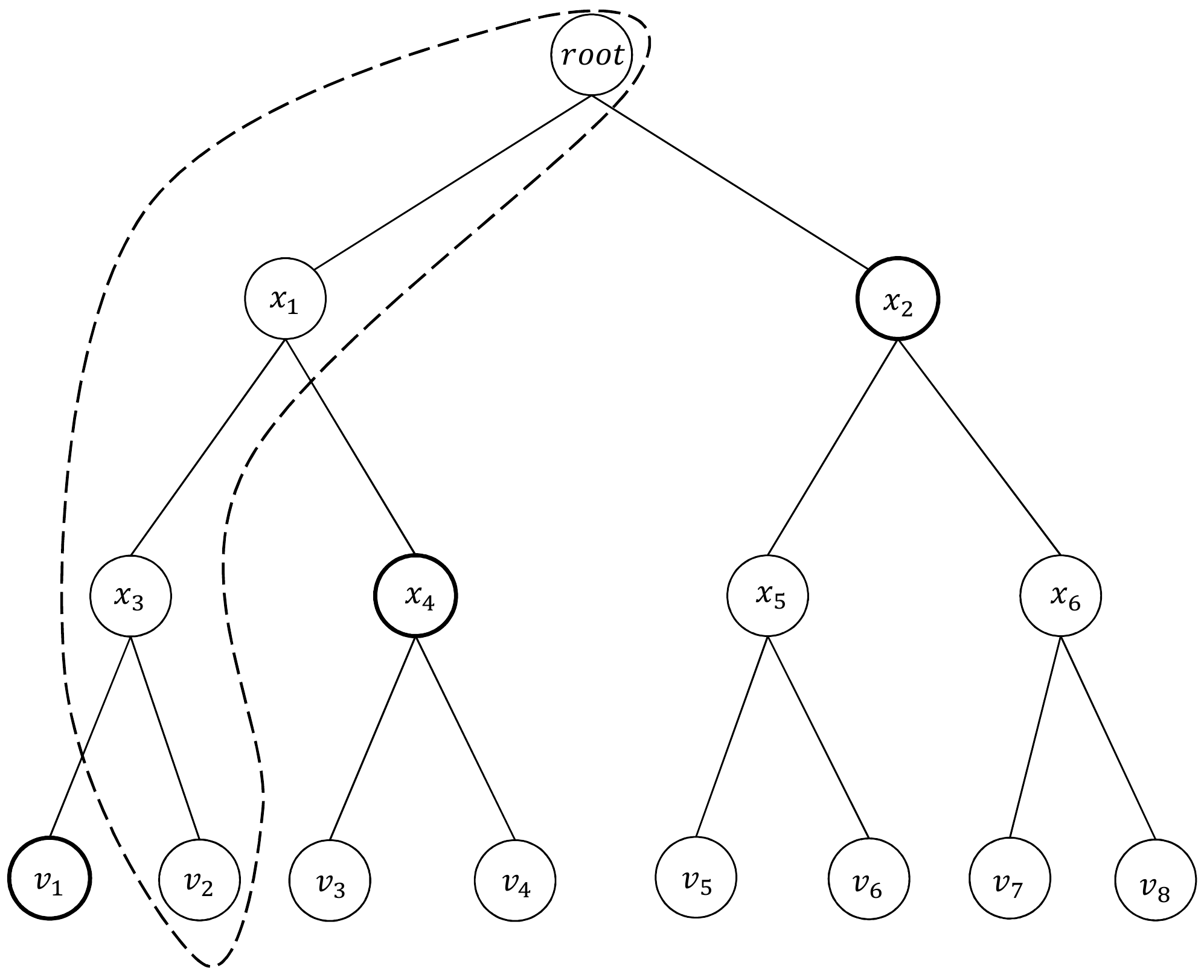}
   \caption{An example of KUNodes algorithm}
   \label{fig:kunodes}
\end{figure}

Figure~\ref{fig:kunodes} gives an example of KUNodes algorithm, where $v_2$ is revoked in this case. Therefore there are three nodes in key updates, which are $v_1$, $x_4$ and $x_2$. The binary tree construction indicates that the size of key updates grows with logarithmic scale, which is better on communication overhead than the original revocation approach (i.e., grows linearly) adopted by related work (for detailed analysis and comparison, see Section~\ref{VI}).

\section{System architecture}\label{III}

In this section, we describe the system architecture and security assumptions of the proposed ring signature-based framework for VANETs.

\subsection{System architecture}\label{III-A}

Generally, our ring signature-based framework consists of four main entities: the Transportation Regulation Center (TRC), Law Enforcement Authority (LEA), RSUs and vehicles equipped with OBUs.
The explanations of these entities are listed as follows.

\begin{itemize}
    \item \textbf{TRC:} TRC is a fully trusted party in the VANETs system with sufficient computation and storage capabilities. As the administrator in VANETs, TRC takes charge of system initialization and registration of the nodes in the network. When a vehicle is misbehaving, TRC also plays a role in pseudonym revocation. These identities of revoked vehicles are recorded in a revocation list named $RL$ by TRC. We assume that TRC can establish a secure channel with RSUs so that RSUs can fetch a fresh revocation list and key updates from TRC confidentially.
    \item \textbf{LEA:} LEA is the agency to reveal the pseudonyms of misbehaving vehicles. In other words, LEA is responsible for detecting fraudulent activities or misconduct of vehicles in VANETs. For instance, if a vehicle broadcasts forged messages anonymously on purpose, then LEA can extract the true identity of the vehicle with the help of TRC.
    \item \textbf{RSUs:} An RSU usually plays an auxiliary role between TRC and vehicles. Namely, RSUs can communicate with TRC through wired or wireless networks and broadcast messages to vehicles in a restricted region. In the proposed framework, the RSUs in the same region have the same regional key pairs. All registered RSUs can obtain fresh node update information from TRC periodically and deliver ring lists to vehicles. 
    \item \textbf{Vehicles:} Each vehicle in this framework is equipped with a communication device called OBU. An OBU contains the hardware security module (HSM), and related cryptographic operations are predefined inside HSM. The HSM can be regarded as a black box, and we assume that the cryptographic operations are always executed correctly by HSM.
\end{itemize}

In the proposed framework, each vehicle has a unique identity (typically denoted by $\mathsf{VID}$). After the initialization of the system by TRC, each vehicle can register itself to obtain a pseudonym and a corresponding private key (typically denoted by $\mathsf{PID}$ and $\mathsf{PSK}$ respectively). Similarly, a regional RSU can obtain a regional identity $\mathsf{RID}$ and a corresponding private key $\mathsf{RSK}$ from TRC. LEA has a private key $s_\mathsf{trac}$ for pseudonym revocation. As we described above, the RSUs periodically request key-update information from TRC. According to the key-update information, RSUs can reject requests from revoked vehicles. The difference between $\mathsf{VID}$ and $\mathsf{RID}$ is that $\mathsf{VID}$ is always kept secret by TRC and vehicles while $\mathsf{RID}$ is always public to vehicles.

When a vehicle enters a certain region, it can obtain a ring list from local RSUs. Since the local $\mathsf{RID}$ is public in VANETs, the vehicle can deliver its pseudonym to the local RSU in a confidential manner, for instance, using utilized identity-based encryption.

Once the local RSU receives the request from a vehicle, it first checks the validity of $\mathsf{PID}$ for the current period. If the $\mathsf{PID}$ is valid, then a ring list containing a set of pseudonyms is returned. When the vehicle obtains the ring list, it can adopt identity-based ring signature scheme to sign the given message. Due to the anonymity property of ring signature, it is not necessary to change vehicle's $\mathsf{PID}$ much more frequently than using ordinary signature. To achieve auditability, a traceable tag is attached to the broadcast message. To decrease the size of key updates, we integrate KUNodes algorithm into our framework.

\subsection{Framework}\label{III-B}

According to the abstract pseudonym life cycle introduced in~\cite{petit2014pseudonym}, we divide our proposed scheme into the following algorithms:

\begin{itemize}
  \item $\mathsf{Setup}(1^\kappa)\to(\mathsf{PP},s,RL,\mathsf{BT})$. Taking the security parameter $\kappa$ as input, this algorithm outputs the public parameter $\mathsf{PP}$, the master secret key $s$, an empty initial revocation list $RL$ and a binary tree $\mathsf{BT}$.
  \item $\mathsf{KeyGen}_s(\mathsf{PP},\mathsf{VID})\to(\mathsf{PID},\mathsf{PSK},\mathsf{Path}(\mathsf{PID}))$. Taking the public parameter $\mathsf{PP}$ and a vehicle's identity $\mathsf{VID}$ as input, this algorithm outputs a key pair ($\mathsf{PID},\mathsf{PSK}$) and the path from $\mathsf{PID}$ to $\mathsf{root}$. Note that the procedure of generating $\mathsf{RID}$ is similar, and we omit the specific description here to avoid redundancy.
  \item $\mathsf{KUNodes}(\mathsf{BT},RL,T)\to ku_{T}$. Taking the binary tree $\mathsf{BT}$, the revocation list $RL$ and current time period $T$ as input, this algorithm outputs the key-update information $ku_{T}$.
  \item $\mathsf{RingReq}_{\mathsf{PSK}}(\mathsf{RID}, \mathsf{PID}, \mathsf{Path}(\mathsf{PID}))\to(C_1,C_2)$. Taking a local RSU's identity $\mathsf{RID}$, the vehicle's pseudonym $\mathsf{PID}$ and the $\mathsf{Path}(\mathsf{PID})$ as input, this algorithm outputs the ciphertext of $\mathsf{PID}$ (denoted by $C_1$) and the ciphertext of $\mathsf{Path}(\mathsf{PID})$ (denoted by $C_2$).  
  \item $\mathsf{RingGen}_{\mathsf{RSK}}(C_1,C_2)\to L/\perp$. Taking two ciphertexts $C_1,C_2$ as input, this algorithm outputs a ring list $L$ or aborts $\perp$.
  \item $\mathsf{Sign}_{\mathsf{PSK}}(m||tag||t,L_s)\to \sigma$. Taking a given message $m$, a tracing tag $tag$, the current timestamp $t$ and a ring list $L_s$ as input, this algorithm outputs a signature $\sigma$.
  \item $\mathsf{Verify}(m||tag||t,L_s,\sigma)\to \mathsf{true}/ \mathsf{false}$. Taking the received message package $m||tag||t$, the ring list $L_s$ and signature $\sigma$ as input, this algorithm returns true or false.
  \item $\mathsf{BatchVer}(M_{batch},L_{batch},\sigma_{batch})\to \mathsf{true}/ \mathsf{false}$. Taking a set of messages $M_{batch}=\{m_1||tag_1||t_1, m_2||tag_2||t_2, \ldots, m_\eta||tag_\eta||t_\eta\}$ and corresponding $L_{batch}=\{L_1, L_2, \ldots, L_\eta\}$, $\sigma_{batch}=\{\sigma_1, \sigma_2, \ldots, \sigma_{\eta}\}$ as input, this algorithm returns true or false.
  \item $\mathsf{Trace}_{s_{\mathsf{trac}}}(tag,L_s)\to \mathsf{PID}$. Taking a tracing tag $tag$ and the corresponding ring list $L_s$ as input, this algorithm outputs the signer's pseudonym $\mathsf{PID}$.
  \item $\mathsf{Revoke}(\mathsf{VID},T,RL)\to RL$. Taking a vehicle's identity $\mathsf{VID}$, the current time period $T$ and revocation list $RL$ as input, this algorithm outputs a fresh $RL$.
\end{itemize}

\subsection{Assumptions of HSM}\label{III-C}

\begin{figure}[htbp]
    \centering
    \includegraphics[width=\linewidth]{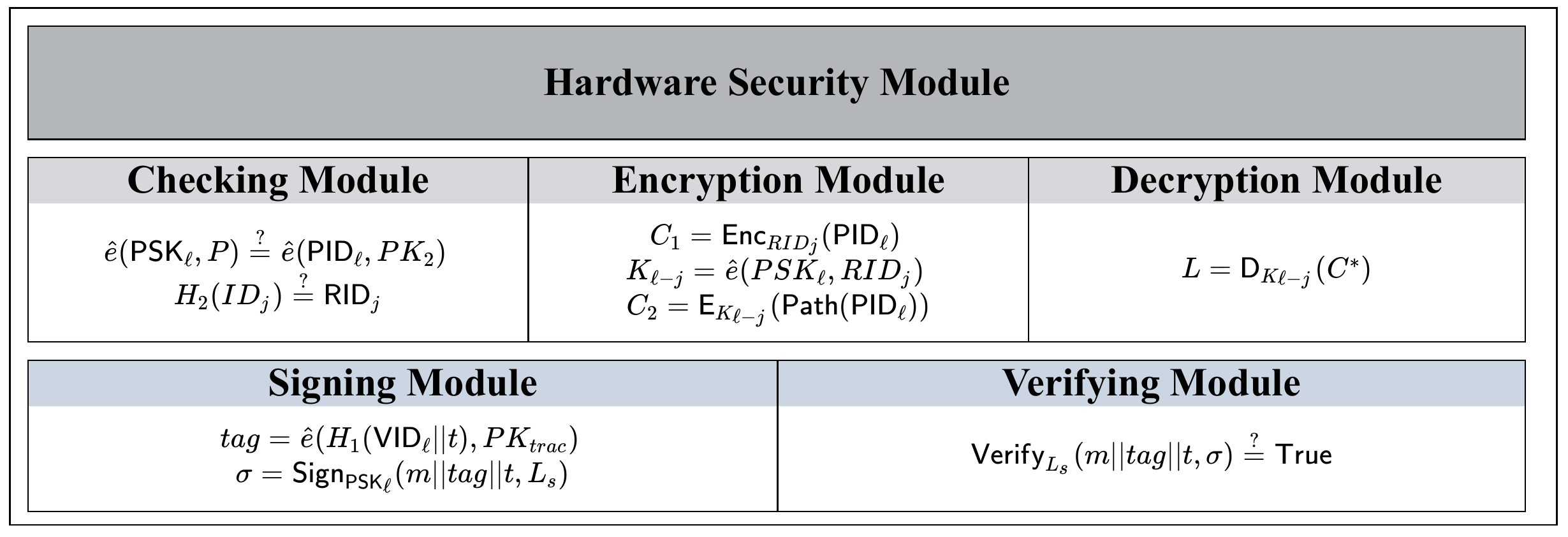}
    \caption{Abstract construction of HSM}
    \label{fig:04}
\end{figure}

Later we will show that the proposed scheme satisfies authentication, conditional anonymity, and auditability. Before discussing these requirements, we recall that each OBU has a hardware called HSM, so that HSM provides an independent environment to perform related cryptographic operations. Each HSM consists of 5 sub-modules as shown in Figure~\ref{fig:04}. Hereafter, all our analysis is based on the assumption of HSMs.

In the case of our proposed scheme, authentication means both sender authentication and message authentication: the message sender must be one member of the ring and the transmitted data cannot be modified. Anonymity guarantees that the receiver only knows the signer is one member of the ring, but cannot determine which exact pseudonym belongs to the signer. As for auditability, there are two aspects: On the one hand, it is required that the true identity can be resolved once the misbehaving vehicle is detected by LEA; On the other hand, the vehicle revocation is achieved by TRC and LEA which assures that only non-revoked vehicles can form a valid ring list from RSU.

\section{The proposed scheme}\label{IV}

\begin{figure}[htbp]
    \centering
    \includegraphics[width=\linewidth]{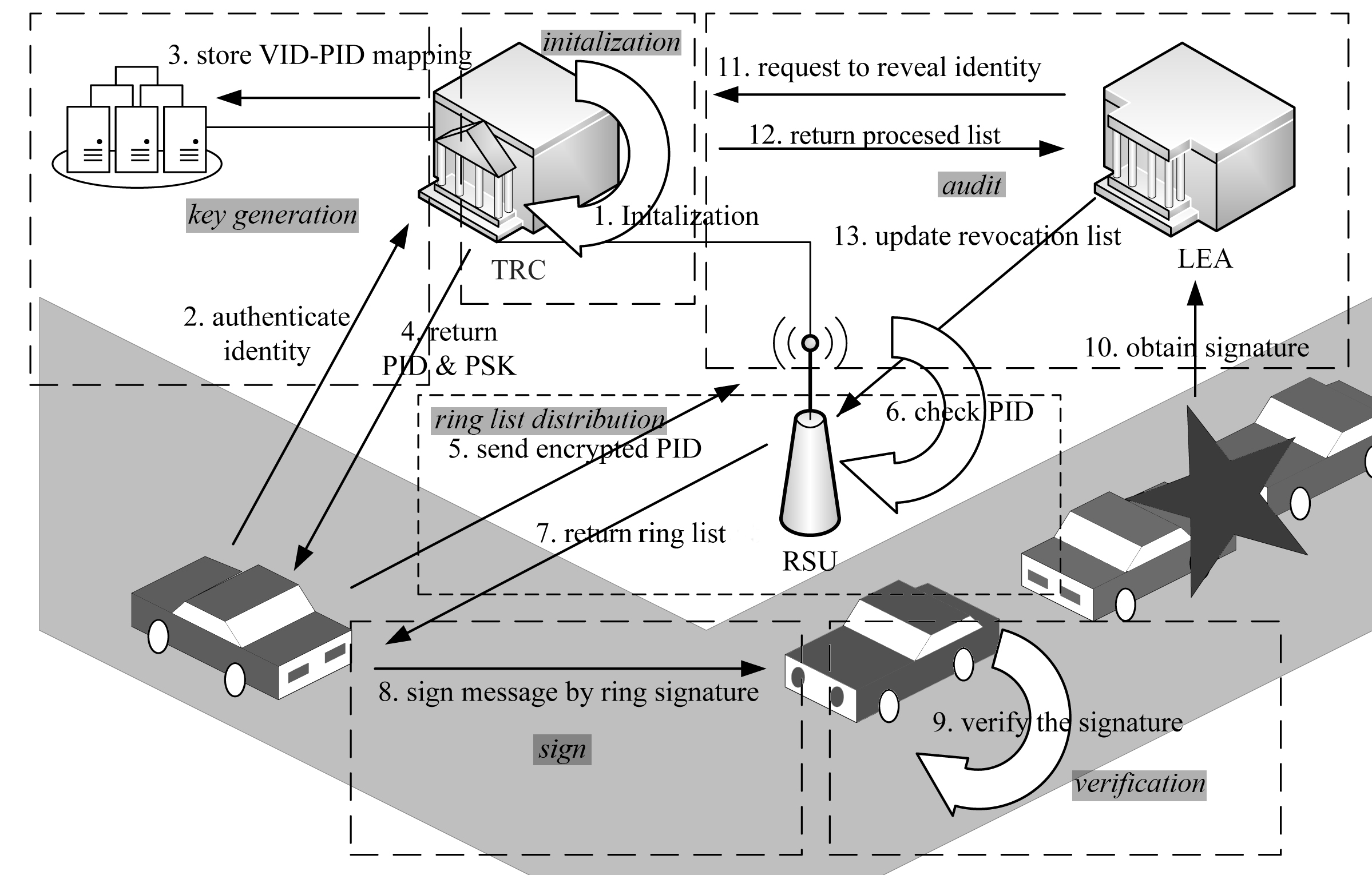}
    \caption{Our proposed scheme}
    \label{fig:03}
\end{figure}

In this section, we use an abstract pseudonym life cycle~\cite{petit2014pseudonym} as shown in Figure~\ref{fig:03} to describe how our proposed scheme works. The whole life cycle can be divided into six phases: {\em initialization}, {\em key generation}, {\em ring list distribution}, {\em sign}, {\em verification} and {\em audit}.
Relevant notations are listed in Table~\ref{tab:01}.

\subsection{Initialization}

TRC first generates a bilinear pairing $\hat{e}:\mathbb{G}_1\times\mathbb{G}_2\to \mathbb{G}_T$ for the preset security parameter $1^\kappa$. Then it chooses the master secret key $s\gets_R\mathbb{Z}_q^*$ randomly, and outputs $\{\mathbb{G}_1,\mathbb{G}_2,\mathbb{G}_T,P,Q,PK_1,PK_2,q,\hat{e},H_1,H_2,H\}$ (see Section~\ref{II-C}). After that, LEA chooses its private key $s_\mathsf{trac}\gets_R\mathbb{Z}_q^*$ randomly. The corresponding public key can be calculated as $PK_\mathsf{trac}=s_\mathsf{trac}\cdot Q$. Finally, TRC outputs the public parameters $\mathsf{PP}=\{\mathbb{G}_1,\mathbb{G}_2,\mathbb{G}_T,P,Q,PK_1,PK_2,PK_\mathsf{trac},q,\hat{e},H_1,H_2,H\}$.

\begin{table}[htbp]
\caption{Notation declarations}
\begin{center}
\begin{tabular}{cc}
\toprule
\textbf{Notations}& \textbf{Explanation}\\
\midrule
$s$ & The master secret key \\
$s_{\mathsf{trac}}$ & The private key of LEA \\
$PK_{\mathsf{trac}}$ & The public key of LEA \\
$\mathsf{PP}$ & Public parameters \\
$\mathsf{VID}_i$& Real ID of vehicle $i$ \\
$\mathsf{PID}_i$& Public key (pseudonym) of vehicle $i$ \\
$\mathsf{PSK}_i$& Private key of vehicle $i$ \\
$\mathsf{RID}_j$& Public key of RSU $j$\\
$\mathsf{RSK}_j$& Private key of RSU $j$\\
$K_{i-j}$ & A shared secret key between $i$ and $j$ \\
$L$ & The ring list generated by RSUs \\
$t_d$ & The expired date of $L$ \\
$L_s$ & The ring list used in ring signature\\
$t$ & The timestamp for signature \\
$tag$ & The traceable tag for signature \\
$\mathsf{KeyGen}(\cdot)$ & The key generation algorithm in Section~\ref{II-C} \\
$\mathsf{Sign}_{sk}(\cdot)$ & The ring signature algorithm in Section~\ref{II-C}\\
$\mathsf{Verify}_{L_s}(\cdot)$ & The verification algorithm in Section~\ref{II-C}\\
$\mathsf{Enc}_{pk}(\cdot)$ & The public encryption algorithm in Section~\ref{II-C}\\
$\mathsf{Dec}_{sk}(\cdot)$ & The public decryption algorithm in Section~\ref{II-C}\\
$\mathsf{E}_{k}(\cdot)$ & A symmetric encryption algorithm (e.g., AES) \\
$\mathsf{D}_{k}(\cdot)$ & A symmetric decryption algorithm corresponding to $\mathsf{E}$\\
$\mathsf{MAC}_{k}(\cdot)$ & A symmetric hash-based message authentication code \\
$a||b$ & String concatenation of a and b \\
$\mathsf{len}(\cdot)$& A function of the number of items in an object \\
\bottomrule
\end{tabular}
\label{tab:01}
\end{center}
\end{table}

\subsection{Key generation}

Before the stage of key generation, TRC first generates an initial binary tree $\mathsf{BT}$ to record vehicles' pseudonyms. After that, for a vehicle with its real identity $\mathsf{VID}_i$, TRC invokes $\mathsf{KeyGen}(\mathsf{VID}_i)$ and sends $(\mathsf{PID}_i,\mathsf{Path}(\mathsf{PID}_i),\mathsf{PSK}_i,\mathsf{PP})$ to the vehicle. Note that these parameters are preloaded into the tamper-proof device HSM. For an RSU with a region identifier $\mathsf{ID}_j$, it can obtain the key pair $(\mathsf{RID}_j, \mathsf{RSK}_j)$ from TRC in the same manner. The specific procedure is given in Algorithm~\ref{alg:keygen}.

\begin{algorithm}
  \begin{algorithmic}[htbp]
  \If {$\mathsf{ID}$ belongs to OBU}
    \State $\mathsf{VID}:=\mathsf{ID}$
    \State $\mathsf{PID}=H_1(\mathsf{VID})$
    \State $\mathsf{PSK}=s\cdot \mathsf{PID}$
    \State assign $\mathsf{PID}$ to an empty leaf node of $\mathsf{BT}$, record $\{\mathsf{PID}:\mathsf{VID}\}$
    \State return $\{\mathsf{PID},\mathsf{Path}(\mathsf{PID}),\mathsf{PSK},\mathsf{PP}\}$
  \ElsIf {$\mathsf{ID}$ belongs to RSU}
    \State $\mathsf{RID}=H_2(\mathsf{ID})$
    \State $\mathsf{RSK}=s\cdot \mathsf{RID}$
    \State return $\{\mathsf{RID},\mathsf{RSK},\mathsf{PP}\}$
  \EndIf
 \caption{$\mathsf{KeyGen}_{s}(\mathsf{ID}):$}\label{alg:keygen}
\end{algorithmic}
\end{algorithm}

\subsection{Ring list distribution}

Once a vehicle, e.g., $V_i$ enters into a certain region, it will obtain a fresh pseudonym list by connecting to the local RSU. In our scheme, the region identifiers are public to all vehicles in VANETs. Therefore, vehicle $V_i$ can deliver its pseudonym to a local RSU in a confidential manner by employing $\mathsf{Enc}_{pk}(\cdot)$ and $\mathsf{Dec}_{sk}(\cdot)$ in Section~\ref{II-C}. The full procedure is described in Algorithm~\ref{alg:acess}. Note that if the bilinear pairing used in our scheme belongs to Type 3, the vehicle's pseudonym will have a shorter representation in an efficient way~\cite{galbraith2008pairings} so that the communication overhead will be decreased.

\begin{algorithm}
  \begin{algorithmic}[htbp]
  \State choose $r\gets_R \mathbb{Z}_q^*$
  \State $g=\hat{e}(PK_1, \mathsf{PID}_j)$
  \State $k_{i-j}=\hat{e}(\mathsf{PSK}_i,\mathsf{RID}_j)$
  \State $C_1:=(rP,\mathsf{PID}_i \oplus H(g^r))$
  \State $C_2=\mathsf{E}_{k_{i-j}}(\mathsf{Path}(\mathsf{PID}_i))$
  \State return ($C_1, C_2$)
 \caption{$\mathsf{RingReq}_{\mathsf{PSK}_i}(\mathsf{RID_j},\mathsf{PID}_i, \mathsf{Path}(\mathsf{PID}_i))$}\label{alg:acess}
\end{algorithmic}
\end{algorithm}

When an RSU receives the request from vehicle $V_i$, it first checks the validity of $V_i$.
If the pseudonym of $V_i$ is valid, then the RSU returns a ring list.
In order to filter the misbehaving vehicles' requests, RSUs need to refresh key-update information from RSU regularly. 
The complete process is given in Algorithm~\ref{alg:ringgen}.
In view of the capability bottleneck of RSUs, we propose that RSUs store the shared keys locally to reduce the computation cost.

\begin{algorithm}
  \begin{algorithmic}[htbp]
  \State parse $C_1$ as $(U,V)$
  \State $\mathsf{PID}_i=V \oplus H(\hat{e}(U,\mathsf{RSK}_j)$
  \If {$\mathsf{Path}(\mathsf{PID}_i)$ $\subseteq$ $\mathsf{KUNodes}(\mathsf{BT},RL,T)$}
    \State choose a ring list $L$, compute $k_{j-i}=\hat{e}(\mathsf{PID}_i, \mathsf{RSK}_j)$
    \State return $(C', \Sigma, t_d)$, where $C' = \mathsf{E}_{k_{j-i}}(L)$, $\Sigma=\mathsf{Mac}_{k_{j-i}}(C'||t_d)$
  \Else
    \State reject the request
  \EndIf
 \caption{$\mathsf{RingGen}_{\mathsf{RSK}_j}(C_1,C_2):$}\label{alg:ringgen}
\end{algorithmic}
\end{algorithm}

\subsection{Sign}
For a vehicle $V_k$ holding a ring list $L$ with an unexpired $t_d$, it first chooses $n'-1$ pseudonyms from $L$ randomly to establish an $n$-length ring list $L_s$, i.e., $L_s=\{\mathsf{PID}_1,\mathsf{PID}_2,\ldots,\mathsf{PID}_k, \ldots,\mathsf{PID}_{n'}\}$. Then it can adopt identity-based ring signature scheme (as described in Section~\ref{II-C}), i.e., $\sigma=\mathsf{Sign}_{\mathsf{PSK}_k}(m||tag||t)$, where $tag=\hat{e}(H_1(\mathsf{VID}_\ell||t), PK_{trac})$. Finally it broadcasts $(m,\sigma,L_s,t,tag)$. The detailed procedure is illustrated in Algorithm~\ref{alg:sign}.

\begin{algorithm}
  \begin{algorithmic}[htbp]
  \State choose $n'-1$ $\mathsf{PID}$ from $L$ randomly
  \State set $L_s:=\{\mathsf{PID}_1,\mathsf{PID}_2,\ldots,\mathsf{PID}_k,\ldots,\mathsf{PID}_{n'}\}$
  \For {$i$ from $1$ to $n'$ $\&\&$ $i\neq k$}
    \State choose $U_i\gets_R \mathbb{G}_1$
    \State $h_i=H(m||tag||t||L_s||U_i)$
  \EndFor
  \State choose $r'\gets_R\mathbb{Z}_q*$
  \State $U_k=r'\mathsf{PID}_k-\sum_{r=1,r\neq k}^{n'}(U_i+h_i\mathsf{PID}_i)$
  \State $h_k=H(m||tag||t||L_s||U_k), V=(h_k+r')\mathsf{PSK}_k$
  \State return $\sigma:=(\{U_i\}_{i=1}^{n'},V)$
 \caption{$\mathsf{Sign}_{\mathsf{PSK}_k}(m||tag||t,L_s):$}\label{alg:sign}
\end{algorithmic}
\end{algorithm}

\subsection{Verification}

When a vehicle, e.g., $V_\ell$ receives $(m,\sigma,L_s,t,tag)$, it first checks $t$ to prevent replay attacks. If valid, then it runs $\mathsf{Verify}_{L_s}(m||tag||t,\sigma)$ to check the validity of the signature. As mentioned above, considering that the capacity of OBUs is limited, we suggest adopting batch verification proposed in~\cite{ferrara2009practical} to speed up the process. Algorithms~\ref{alg:veri} and \ref{alg:batch} present the procedures of single verification and batch verification, respectively.

\begin{algorithm}
  \begin{algorithmic}[htbp]
  \State $\sigma=(\{U_i\}_{i=1}^{n'},V)$
  \For {$i$ from $1$ to $n'$}
    \State $h_i=H(m||tag||t||L_s||U_i)$
  \EndFor
  \State return {$\hat{e}(\sum_{i=1}^{n'}(U_i+h_i\mathsf{PID}_i),PK_2)\overset{?}{=}\hat{e}(V,Q)$}
 \caption{$\mathsf{Verify}(m||tag||t,L_s,\sigma):$}\label{alg:veri}
\end{algorithmic}
\end{algorithm}

\begin{algorithm}
  \begin{algorithmic}[htbp]
  \State $\sigma_{batch}=\{\sigma_1,\sigma_2,\ldots,\sigma_\eta\}$
  \State $M_{batch}=\{m_1||tag_1||t_1,m_2||tag_2||t_2,\ldots,m_{\eta}||tag_\eta||t_\eta\}$
  \State $L_{batch}=\{L_{s1},L_{s2},\ldots,L_{s\eta}\}$
  \For {$i$ from $1$ to $\eta$}
    \For {$j$ from $1$ to $\mathsf{len}(L_{si})$}
      \State $h_{ij}=H(M_i||L_{si}||U_{ij})$
    \EndFor
  \EndFor
  \State return $\hat{e}(\sum_{i=1}^{\eta}\sum_{j=1}^{\mathsf{len}(L_{si})}(U_{ij}+h_{ij}\mathsf{PID}_{ij}), PK_2)\overset{?}{=}\hat{e}(\sum_{i=1}^{\eta}V_i,Q)$
 \caption{$\mathsf{BatchVer}(M_{batch},L_{batch},\sigma_{batch}):$}\label{alg:batch}
\end{algorithmic}
\end{algorithm}

In the procedure of batch verification, once the result is false, which means that at least one signature in $\sigma_{batch}$ is invalid, then we can use the \textit{divide-and-conquer technique}~\cite{ferrara2009practical} to exclude the invalid signatures.

\subsection{Tracing and Revocation}

When LEA detects misbehaviors in VANETs, it calculates $tag'=tag^{1/s_{trac}}$ and then sends $\{L_s$,$t\}$ to TRC. For $L_s=\{\mathsf{PID}_1,\mathsf{PID}_2,\ldots,\mathsf{PID}_{n'}\}$, TRC computes $H_i'=\hat{e}(H_i(\mathsf{VID}_i||t),P)$, where $i \in \{1,2,\ldots,n'\}$, and returns $\cup_{i=1}^{n'}\{H_i'\}$. By comparing $tag'$ and $\cup_{i=1}^{n'}\{H_i'\}$, LEA can determine the signer's pseudonym. Furthermore, LEA would find out the signer's true identity by sending the pseudonym to TRC. To revoke a vehicle's identity $\mathsf{VID}$, TRC first takes the current timestamp $T$ and the leaf node $n_v$ associated with $\mathsf{VID}$. Then TRC adds $(n_v, T)$ to $RL$.

\section{Security analysis}\label{V}

\subsection{Correctness}

In V2I communication, when vehicle $\mathsf{V}_\ell$ enters the region within the range of $\mathsf{RSU}_j$, it will receive the broadcasting $\mathsf{RID}_j$ in this region. Once $\mathsf{V}_\ell$ obtains $\mathsf{RID}_j$, it can invoke the Checking Module to check the validity of $\mathsf{RSU}_j$. In the process of delivering $\mathsf{PID}_\ell$, the correctness and security are guaranteed by the property of identity-based encryption scheme~\cite{boneh2001identity}. According to the property of bilinear pairing, we know that:
\begin{equation}
\begin{split}
K_{\ell-j}&=\hat{e}(\mathsf{PSK}_\ell,\mathsf{RID}_j) \\
&=\hat{e}(\mathsf{PID}_\ell,\mathsf{RID}_j)^s  \\
&=\hat{e}(\mathsf{PID}_\ell,\mathsf{RSK}_j) \\
&=K_{j-\ell}.\\
\end{split}
\label{eq:kij}
\end{equation}
It is clear that both $\mathsf{RSU}_j$ and $\mathsf{V}_\ell$ obtain the same shared key, and can establish an efficient trusted channel for further communication through symmetric cryptography.

In V2V communication, if the procedure of signing a message $m$ is executed correctly, then the corresponding signature $\sigma$ must satisfy the verifying equation~(\ref{eq:kij}).

\subsection{Unforgeability}

We call a signature $\sigma$ unforgeable, if an adversary cannot generate a signature for a new message, given a few signatures corresponding to the messages of his own choice. Since the ring signature scheme~\cite{chow2005efficient} we employed has been proven to be unforgeable against chosen message attacks in the random oracle model, we note that our protocol is also unforgeable.

\subsection{Conditional anonymity}

In V2I communication, RSU $j$ only knows the pseudonym of $\mathsf{V}_\ell$ rather than the true identity of $\mathsf{V}_\ell$, i.e.,$\mathsf{VID}_\ell$. As for V2V communication, the true signer is hidden in a set of pseudonyms $L$. For any eavesdroppers in VANETs, they cannot figure out the true signer from $L$ even though they know the corresponding $tag$. Only the LEA can identify the signer in $L$ through the secret tracing key $s_{trac}$.

\subsection{Against replay attacks}

Note that each message contains a timestamp in our scheme. This indicates that once vehicles figure out that a message is expired, then this message will be abandoned before being verified. If an adversary forges a fresh timestamp to replace the original one, then this message must not be able to pass the verification.

\section{Performance analysis}\label{VI}

In this section, we evaluate the performance of the proposed scheme in terms of both computation cost and communication cost. Specifically, we invoke CHARM~\cite{charm13}, a framework for rapidly prototyping advance cryptosystems, to implement the proposed scheme. To compare the impact of different paring settings, we choose two elliptic curves named ``SS512'' and ``MNT159'' in our experiments, where ``SS512'' represents a super singular curve (symmetric bilinear pairing) with a 512-bit base filed in $\mathbb{G}_1$, and ``MNT159'' represents the Miyaji, Nakabayashi, Takano curve (asymmetric bilinear pairing) with a 159-bit base filed in $\mathbb{G}_1$. Table~\ref{tab:02} lists the difference between these two elliptic curves.

\begin{table}[htbp]
\centering
\begin{adjustbox}{max width=0.5\textwidth}
\begin{threeparttable}
\caption{The security level of the utilized elliptic curves\tnote{$\dagger$}}
\begin{tabular}{cccccc}
    \toprule
    Type & $\mathbb{G}_1$ & $\mathbb{G}_2$ & $\mathbb{G}_T$ & Security level & Paring Type \\
    \midrule
    MNT159 & 159 bits & 477 bits & 945 bits & 70 bits & Type 1 \\
    SS512 & 512 bits & 512 bits & 1024 bits & 80 bits & Type 3 \\
    \bottomrule
\end{tabular}
\begin{tablenotes}
    \footnotesize
  \item[$\dagger$] For more information on the security level of elliptic curves and types of bilinear pairings, we refer to~\cite{galbraith2008pairings,lenstra2001selecting}.
\end{tablenotes}
\label{tab:02}
\end{threeparttable}
\end{adjustbox}
\end{table}

To simulate the limited computation capacity of OBUs in VANETs, our experiments are performed on Raspberry Pi 4 Model B, which is a cheap microcomputer with a 1.5 GHz ARM Cortex-A71 CPU and 4 GB RAM running Debian Linux operation system. All tested cases are executed 1000 times and runtime is measured in CPU time of the current process.

\subsection{Computation cost}

First, we list the runtime of each individual operation under ``SS512'' and ``MNT159'' in Table~\ref{tab:runtime}.

\begin{table}[htbp]
\centering
\begin{adjustbox}{max width=0.5\textwidth}
\begin{threeparttable}
\caption{The runtime\tnote{$\dagger$} of executing different cryptographic operations in different Elliptic Curves}
\begin{tabular}{ccccc}
    \toprule
    Operations & MNT159 & SS512 \\
    \midrule
    The execution time of one scale multiplication operation on $\mathbb{G}_1$ & 1.58 & 5.02\\
    The execution time of one map-to-point hash operation on $\mathbb{G}_1$ & 0.16 & 11.67\\
    The execution time of one bilinear pairing operation & 11.34 & 6.47\\
    The execution time of one exponentiation operation on $\mathbb{G}_T$ & 3.47 & 0.79\\
    \bottomrule
\end{tabular}
\begin{tablenotes}
    \footnotesize
  \item[$\dagger$] The execution time is measured in milliseconds, and evaluated by the average for 1000 times execution.
\end{tablenotes}
\label{tab:runtime}
\end{threeparttable}
\end{adjustbox}
\end{table}

These results in Table~\ref{tab:runtime} are consistent with relevant study~\cite{cui2018efficient}, and it can be seen that the operations on $\mathbb{G}_1$ under ``MNT159'' is much more efficient than that under ``SS512'', while the bilinear pairing operation in ``MNT159'' is more time-consuming. This also indicates that the choice of bilinear pairings depends on the specific constructions of utilized methods.

\begin{figure}[htbp]
   \centering
   \includegraphics[width=0.9\linewidth]{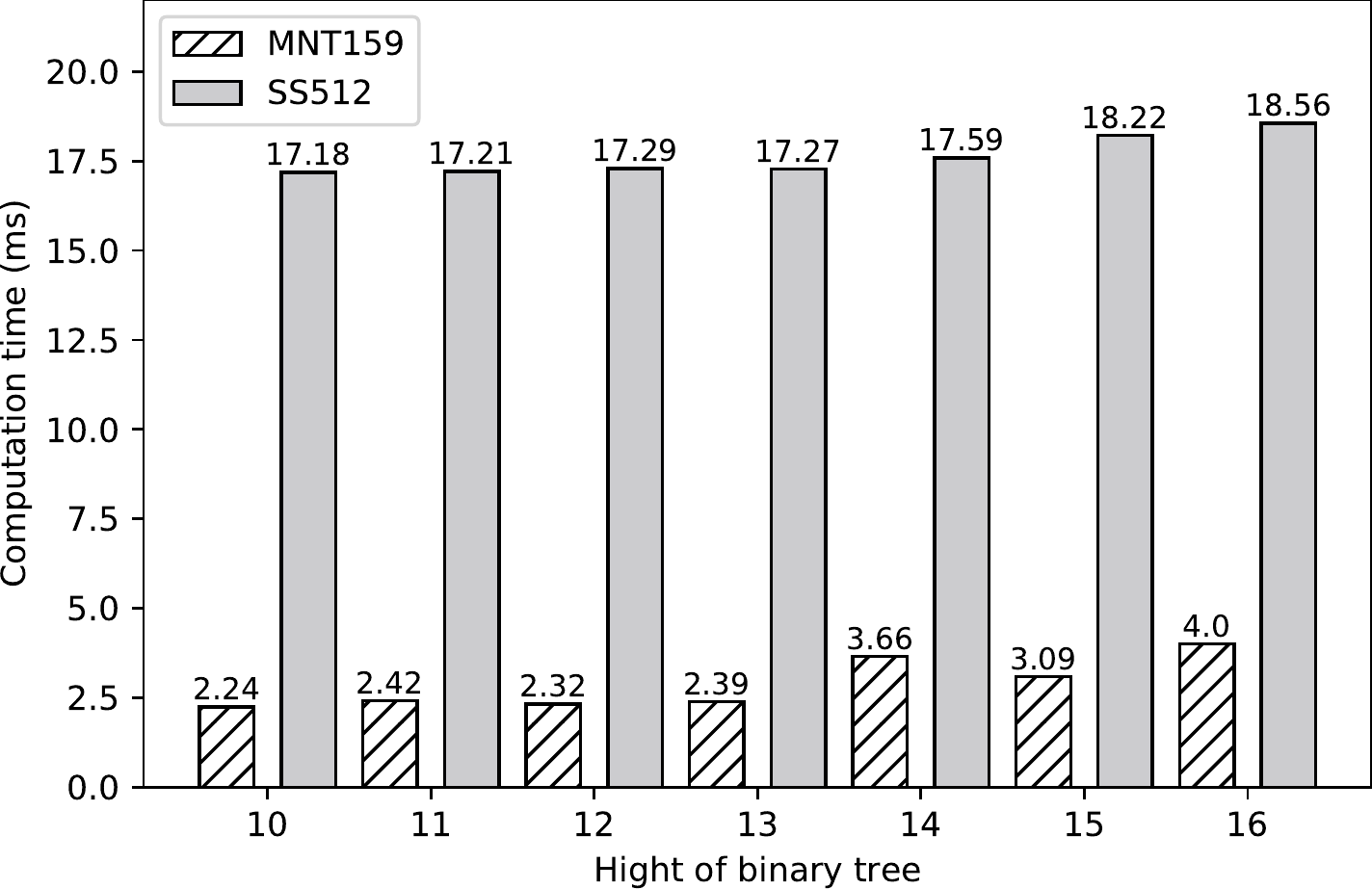}
   \caption{The computation cost for an OBU generation}
   \label{fig:obu_gen}
\end{figure}

For the proposed scheme, we first test the computation cost for TRC in registering an OBU. According to Figure~\ref{fig:obu_gen}, we can observe that the generation time of an OBU increases with the height of $\mathsf{BT}$. Under the same conditions, the operation of OBU generation under MNT159 setting is more efficient than that in SS512 setting. 

\begin{figure*}[htbp]
  \centering
  \subfloat[V2I]{
    \includegraphics[width=0.5\linewidth]{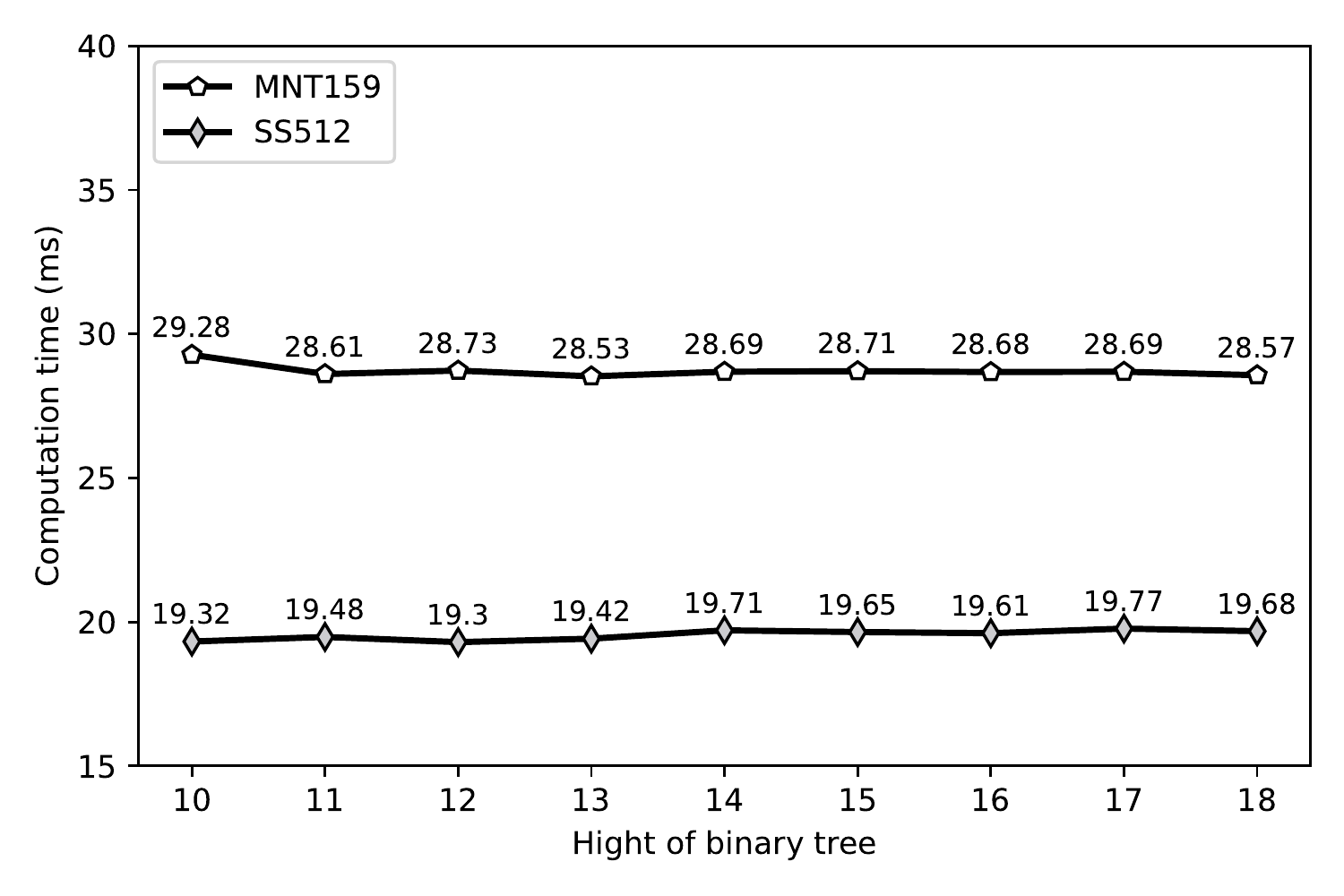}
    \label{fig:V2I}
  }
  \subfloat[I2V]{
    \includegraphics[width=0.5\linewidth]{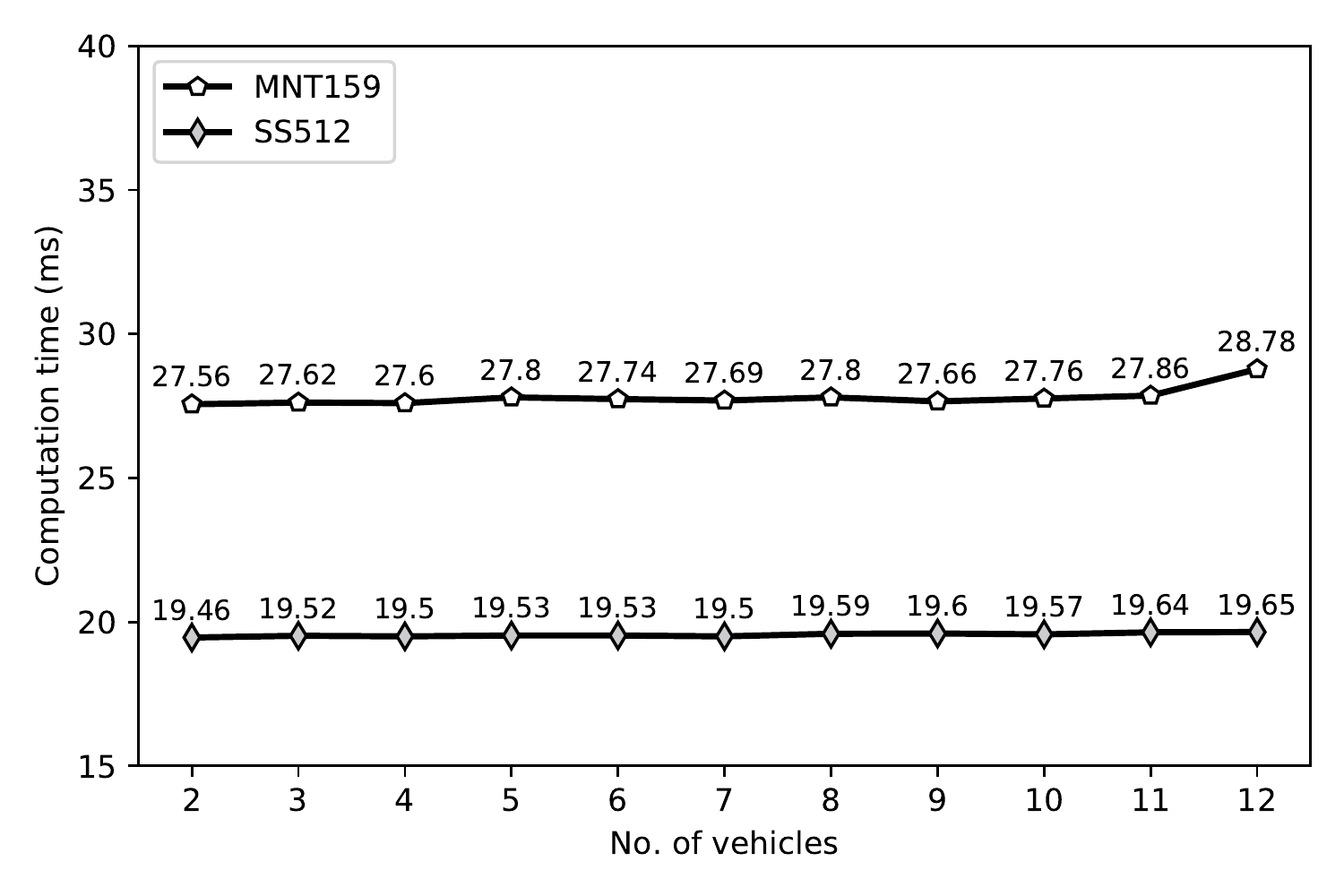}
    \label{fig:I2V} 
  }
  \caption{The computation cost for V2I \& I2V}
  \label{fig:v2i&i2v}
\end{figure*}

\begin{figure*}[htbp]
  \centering 
  \subfloat[Signing]{ 
    \label{fig:sign}
    \includegraphics[width=0.48\textwidth]{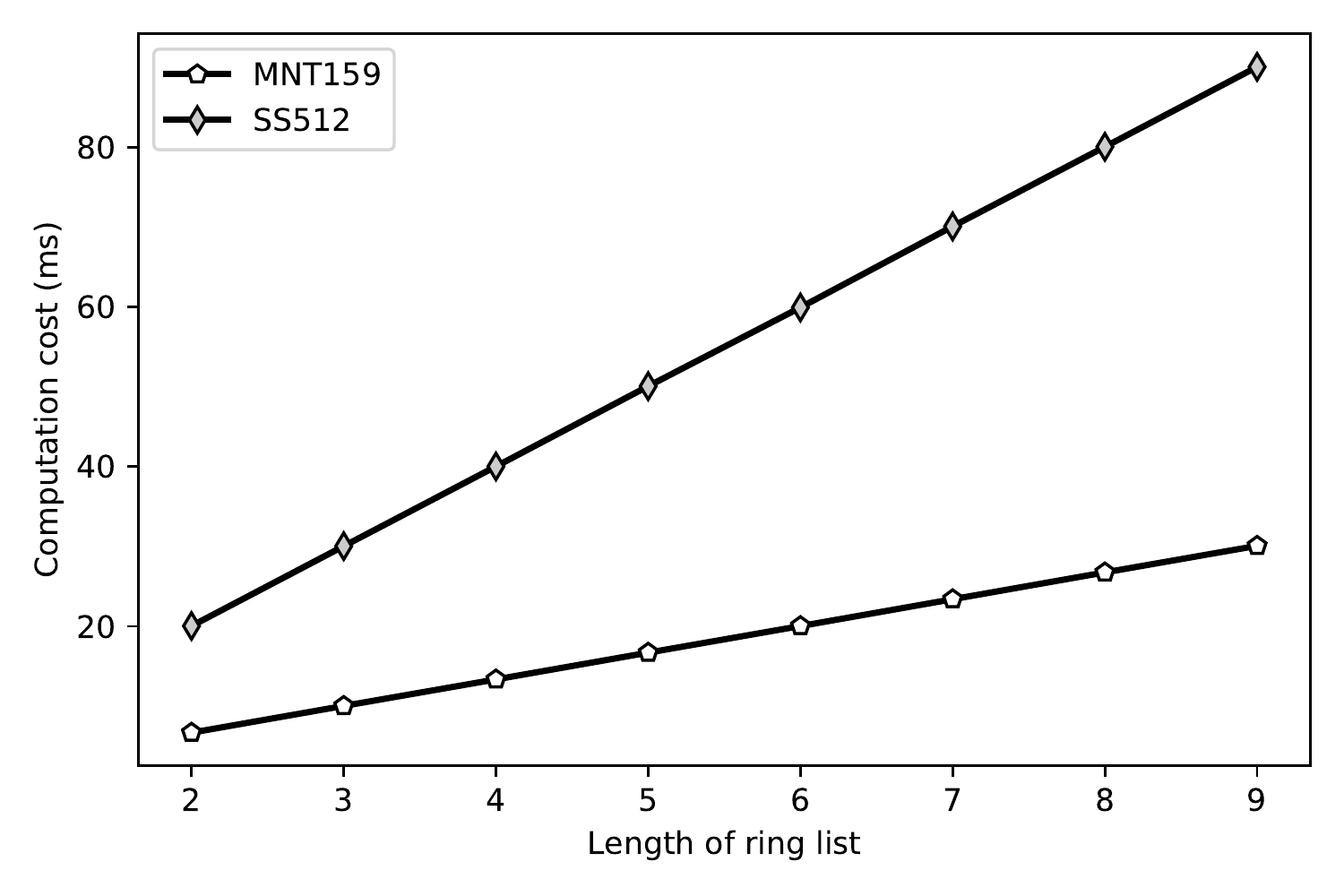}}
  \subfloat[Verification]{ 
    \label{fig:veri} 
    \includegraphics[width=0.48\textwidth]{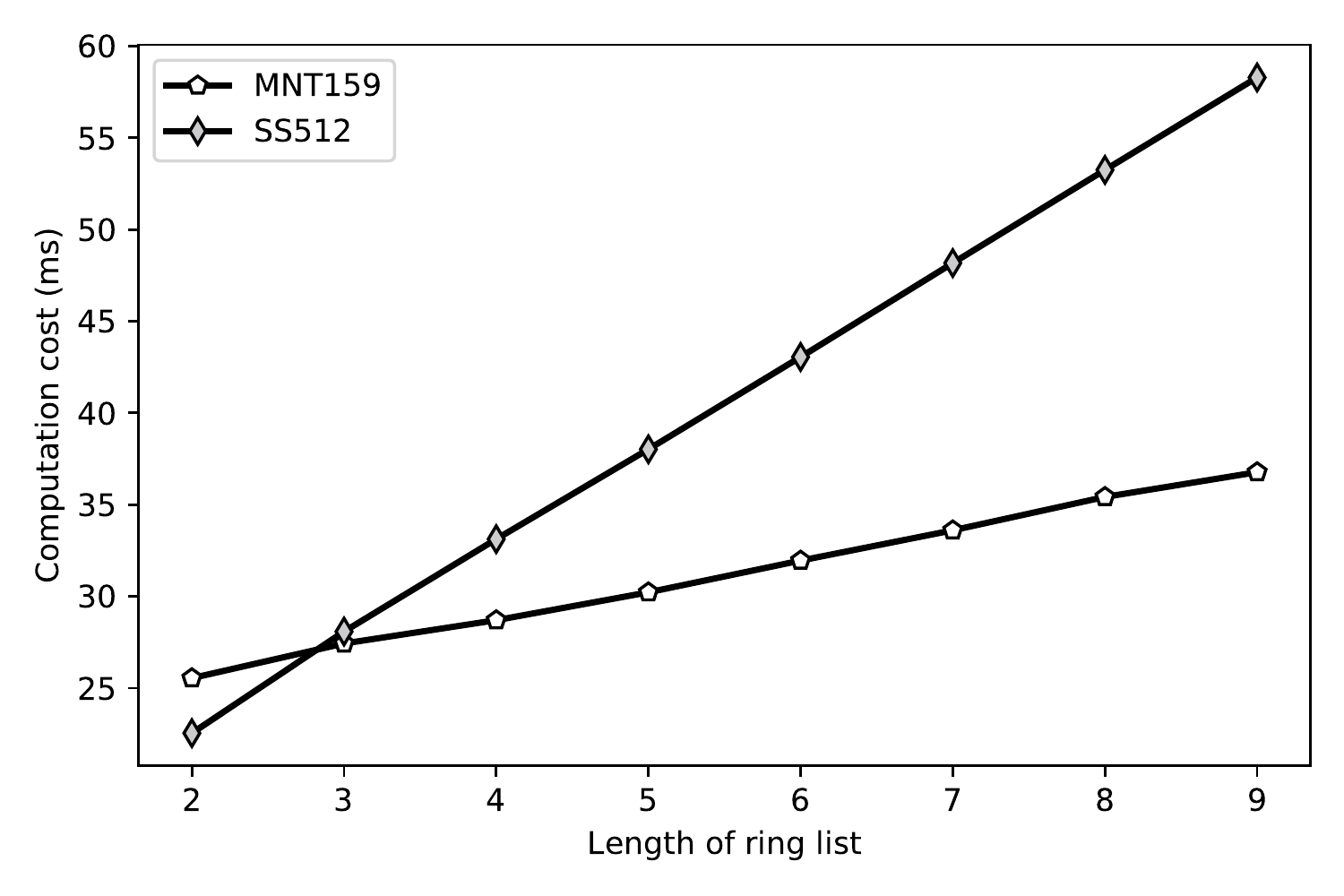}} 
  \caption{The computation cost for signing \& verification}
  \label{fig:sig&veri} 
\end{figure*}

\begin{figure*}[htbp]
  \centering 
  \subfloat[No. of signatures = 10]{ 
    \label{fig:batch1} 
    \includegraphics[width=0.48\linewidth]{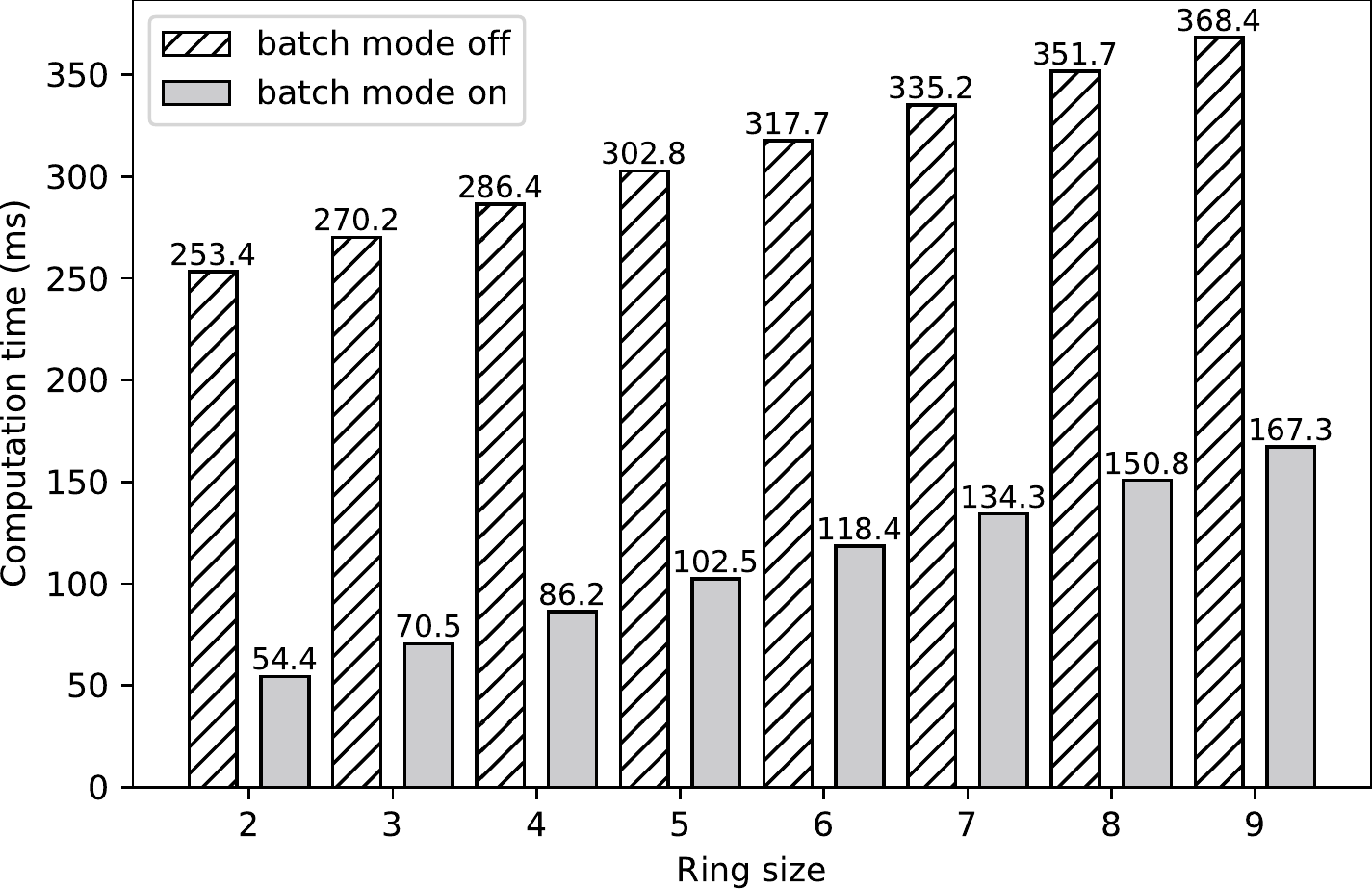}}
  \subfloat[No. of signatures = 20]{
    \label{fig:batch2} 
    \includegraphics[width=0.48\linewidth]{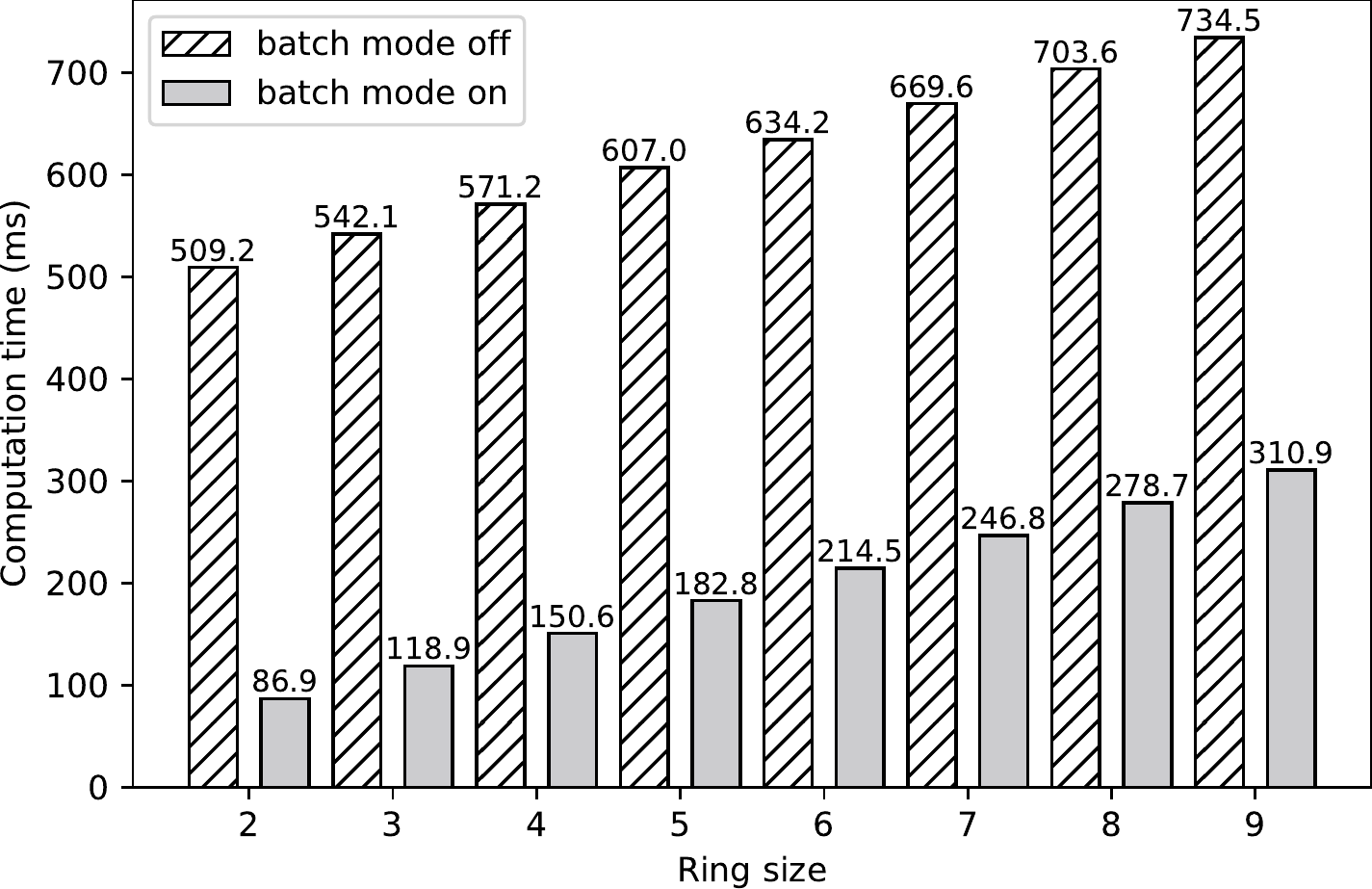}}
  \\
  \subfloat[Difference between batch on and off]{
    \label{fig:batch3} 
    \includegraphics[width=0.48\linewidth]{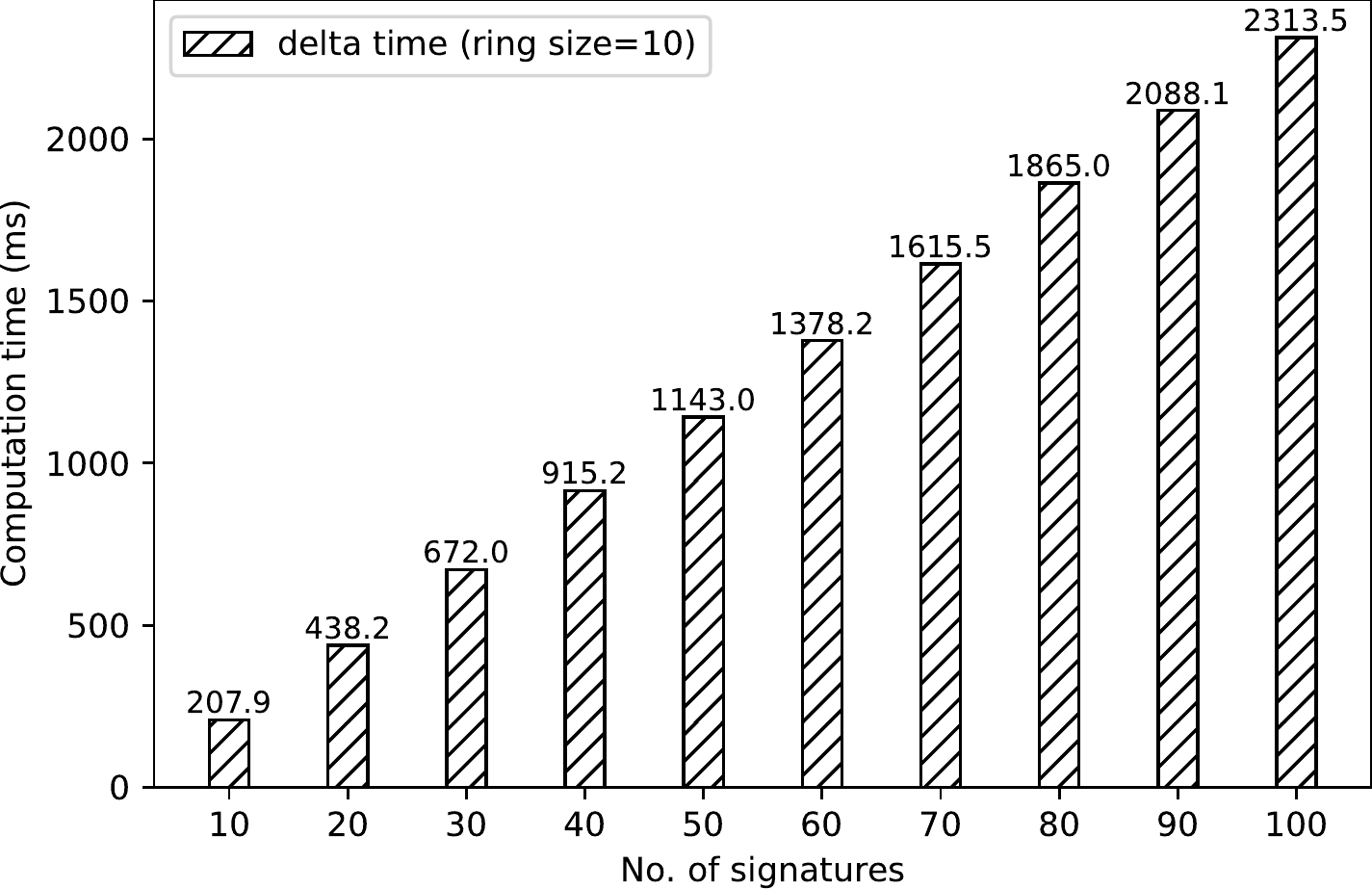}}
  \subfloat[Batch verification under different ring size]{
    \label{fig:batch4} 
    \includegraphics[width=0.48\linewidth]{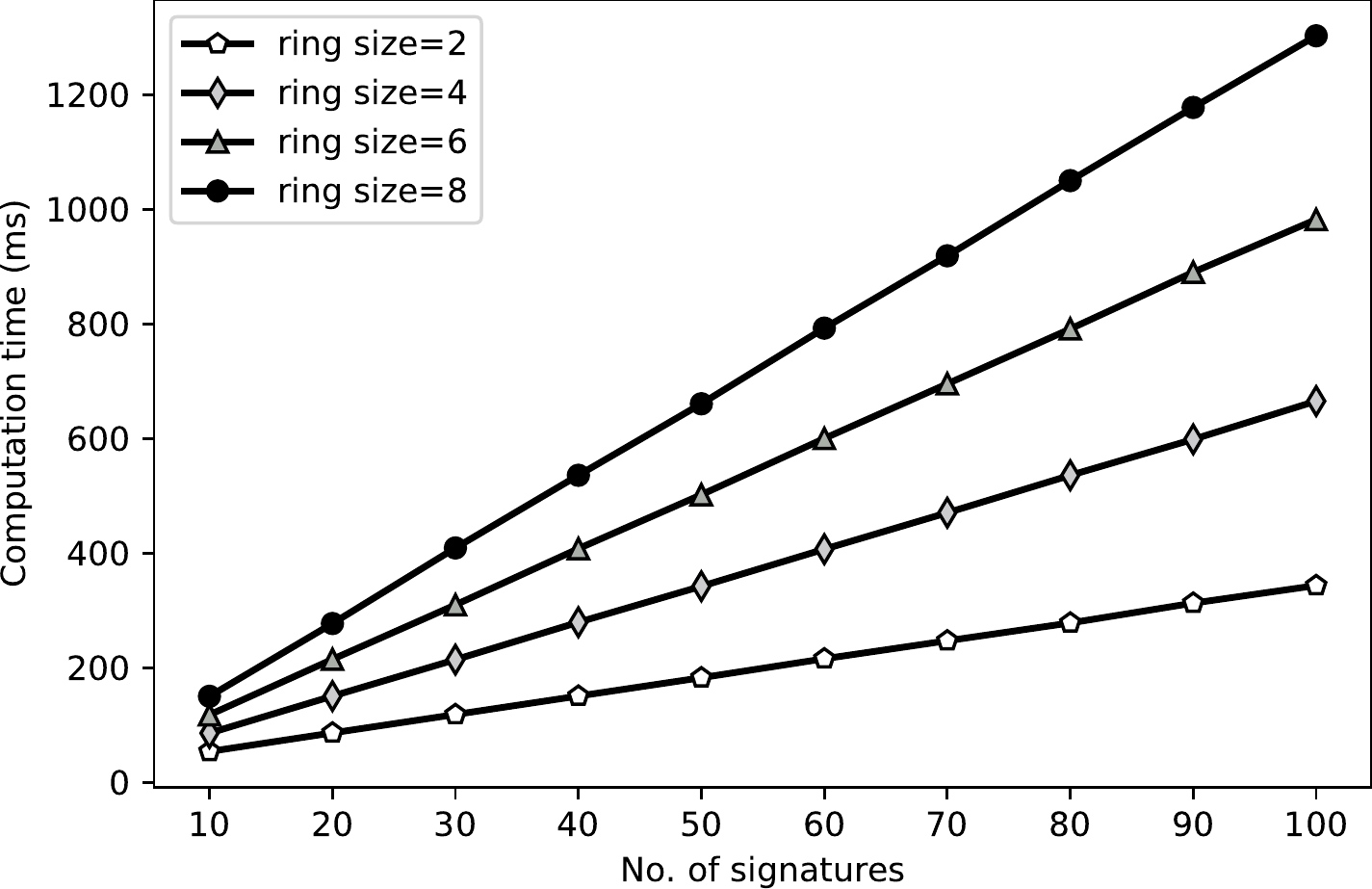}}
  \caption{The computation cost for batch verification}
  \label{fig:batch} 
\end{figure*}

Figure~\ref{fig:v2i&i2v} shows the average computation cost for V2I and I2V phases, respectively.
Note that V2I phase contains two essential encryption methods: one is identity-based encryption for vehicle's pseudonym, and the other is symmetric encryption (we employed AES in our experiments) for vehicles' paths.
According to Figure~\ref{fig:V2I}, it seems that the height of $\mathsf{BT}$ has little effect on vehicle's computational efficiency. The average computation time under ``SS512'' curve is slightly more efficient (by about 10 ms) than that under ``MNT159'' curve. In terms of the computation cost for I2V (see Figure~\ref{fig:I2V}), we set the default height of $\mathsf{BT}$ to $20$, which means the maximum number of vehicles is $2^{20}=1,048,576$. Hereafter, unless otherwise stated, the default height of $\mathsf{BT}$ is set to 20. The result also shows that the number of requests from vehicles has a limited influence on I2V computational efficiency. The computation time under ``SS512'' curve is slightly more efficient (by about 8 ms) than that under ``MNT159'' curve.

Figure~\ref{fig:sig&veri} presents the computation cost for signing and verification, respectively. It is not difficult to see that the computation cost grows linearly as the length of ring list. Since in ring signature, the signer can enhance the level of anonymity by choosing a longer ring list. In other words, vehicles can adjust the size of ring list based on the demand of traffic environment in a flexible way. The results also show that ``MNT159'' has a better performance in signing and verification than ``SS512'', so it is highly recommended that using ``MNT159'' in our proposed scheme for efficiency.

Figure~\ref{fig:batch} shows the impact of batch verification in our scheme. As illustrated in Figure~\ref{fig:batch1}, we choose 10 signatures randomly and record the total CPU time via regular verification and batch verification, respectively. When we enable the batch mode, the procedure of verification can save around 200ms no matter which ring size we used. If we change the number of signatures from 10 to 20, the gap of verification cost
becomes larger (see Figure~\ref{fig:batch2}). Figure~\ref{fig:batch3} also confirms that when the number of signatures increases, the difference between batch verification and regular verification is greater.

There are two important factors which determine the efficiency of batch verification: the number of signatures, and the ring size. In Figure~\ref{fig:batch4}, we focus the influence of these two factors in batch verification. As shown in this figure, there is a trade-off between privacy and efficiency when adopting a longer ring list.

\subsection{Communication cost}

Due to the fact that we adopt pseudonym rather than certificates in the proposed scheme, the proposed scheme have advantages over those schemes based on PKI in communication cost. To illustrate the communication cost of our scheme clearly, we summarize the storage overhead in the following table.

\begin{table}[htbp]
\centering
\begin{adjustbox}{max width=0.5\textwidth}
\begin{threeparttable}
\caption{Communication cost of the proposed scheme\tnote{$\dagger$}}
\begin{tabular}{cccc}
\toprule
  Type & Size of pseudonym & Size of signature & Size of key updates \\
\midrule
MNT159 &  30 bytes  &  $30(|L_s|+1)$ bytes  & \multirow{2}*{$\mathcal{O}(r\log(N/r))$ or $\mathcal{O}(N-r)$} \\
SS512  &  90 bytes  &  $90(|L_s|+1)$ bytes  & \\
\bottomrule
\end{tabular}
\begin{tablenotes}
    \footnotesize
  \item[$\dagger$] The communication cost is evaluated by the built-in function $\mathsf{serialize()}$ with enabling compression in CHARM.
\end{tablenotes}
\label{tab:communicaton}
\end{threeparttable}
\end{adjustbox}
\end{table}

In Table.~\ref{tab:communicaton}, we list the size of pseudonym, size of signature and size of key updates respectively in different pairing settings. Let $|L_s|$ represent the length of ring list used in ring signature, and the size of a signature is linear to the size of ring list. Along with the CHARM in our experiments, the size of an element in $\mathbb{G}_1$ is about $30$ bytes in ``MNT159'' and $90$ bytes in ``SS512'' after serialization. According to Algorithm~\ref{alg:sign}, a signature on a message consists of two parts, namely $\sigma:=({U_i}_{i=1}^{n'}, V)$, where $U_i$ and $V$ both belong to $\mathbb{G}_1$.

\begin{figure}[htbp]
   \centering
   \includegraphics[width=\linewidth]{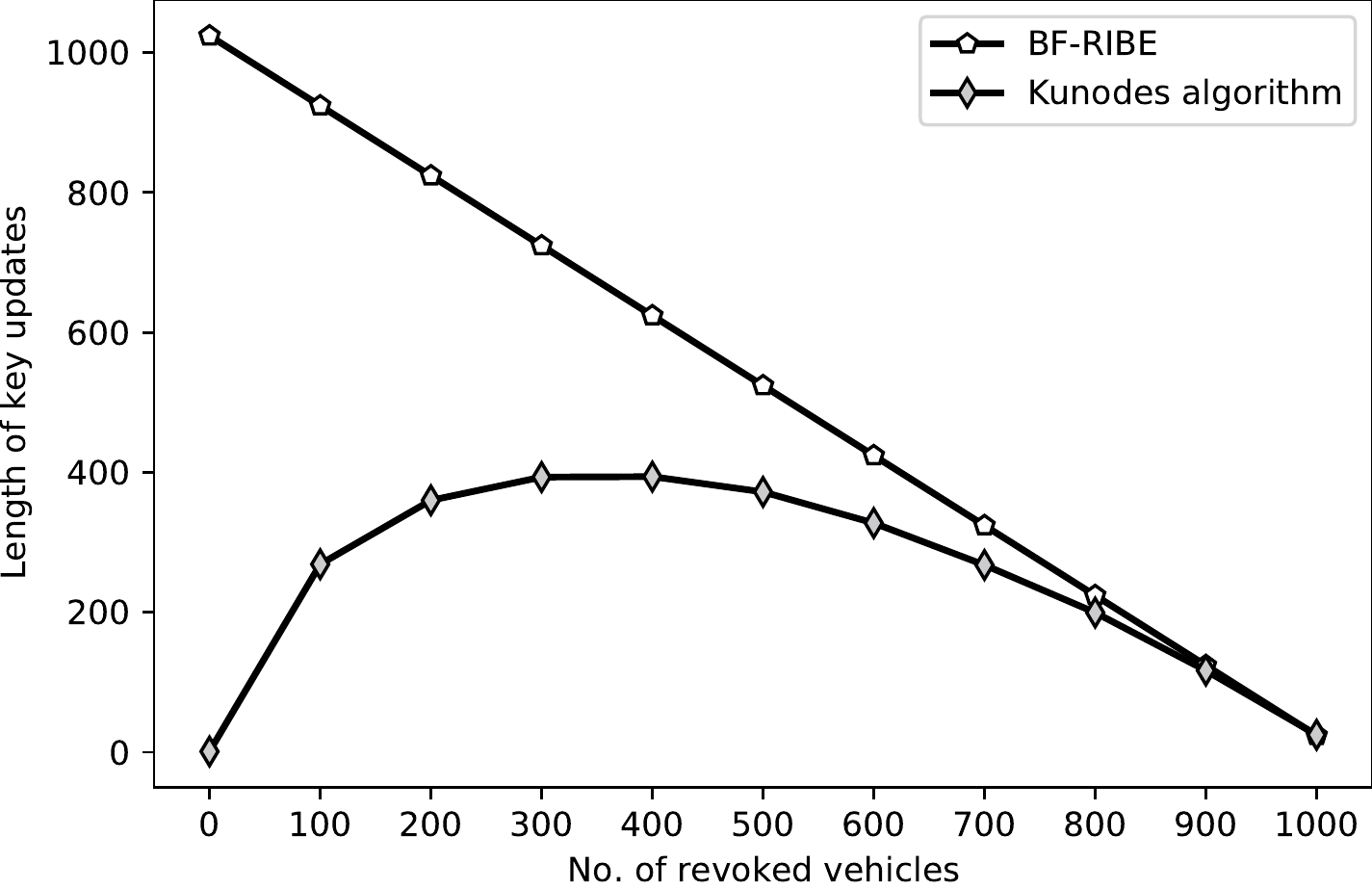}
   \caption{Communication overhead for key updates}
   \label{fig:keyup}
\end{figure}

As for the size of key updates, we denote $N$ and $r$ as the number of all vehicles and revoked vehicles, respectively. Similar to other applications (e.g.~\cite{cui2019attribute,roca2016revocation}) employing the KUNodes algorithm, when $1< r \leq \frac{N}{2}$, the communication cost of key updates is roughly $\mathcal{O}(r\log(N/r))$; when $\frac{N}{2} < r \leq N$, the communication cost of key updates is roughly $\mathcal{O}(N-r)$.

In Figure~\ref{fig:keyup}, we set $N=2^{10}=1,024$ and measured the number of key updates by employing KUNodes algorithm and Boneh-Franklin original revocable IBE~\cite{boneh2001identity} (BF-RIBE for short), respectively. As we mentioned in Section~\ref{relatedwork}, several schemes based on ring signature have not paid enough attention to revocation or just applied the approach in BF-RIBE. Figure~\ref{fig:keyup} shows that the KUNodes algorithm is much more efficient than BF-RIBE especially when $1< r \leq \frac{N}{2}$.

\section{Conclusion}\label{VII}

In this paper, we propose an efficient identity-based batch verification scheme for VANETs based on ring signature. Unlike other ring signature-based schemes, we restrict the generation of a ring to avoid disruptions from malicious vehicles. Considering that VANETs are usually highly dense in most real-world scenarios, we adopt batch verification to reduce the computation cost. In terms of the communication overhead, we compare two different types of bilinear
pairings and show that Type 3 bilinear pairings have a shorter size which is more suitable in the proposed scheme. Besides, we integrate KUNodes algorithm in the key-update phase to decrease the communication cost.  To simulate the environment of OBUs, we implement the proposed scheme on the Raspberry Pi 4 Model B platform. The results also show that our scheme is much more efficient in both computation and communication cost in batch mode.

As a possible direction of future work, it might be interesting to consider building HSMs in real-world applications based on the trusted execution environment (TEE) such as ARM's TrustZone.

\ifCLASSOPTIONcaptionsoff
  \newpage
\fi




\end{document}